\documentclass[english,american,superscriptaddress,prd,a4paper,nofootinbib,11pt,eqsecnum,secnumarabic]{revtex4-1}
\usepackage{multirow,slashed,float,graphicx,caption,subcaption,hyperref,color}
\usepackage[dvipsnames]{xcolor}
\usepackage[utf8]{inputenc}
\hypersetup{colorlinks=true, linkcolor=BurntOrange, citecolor=BurntOrange,filecolor=BurntOrange, urlcolor=BurntOrange}

\newcommand{\be}{\begin{equation}}
\newcommand{\ee}{\end{equation}}
\newcommand{\bea}{\begin{eqnarray}}
\newcommand{\eea}{\end{eqnarray}}

\def\bsp#1\esp{\begin{split}#1\end{split}}

\usepackage{epsfig}
\usepackage{amsbsy}
\usepackage{varioref}
\usepackage{pifont}
\usepackage{amsmath,amssymb}
\usepackage{wrapfig}

\usepackage[normalem]{ulem}

\usepackage{mdframed}
\usepackage[export]{adjustbox}

\begin{document}

%%%%%%%%%%%%%%%%%%%%%%%%%%%%%%%%%%%%%%%%%%%%%%%%%%%%%%%%%%%%%%%%%%%%%%%%%%%
\title{Vector-Like top quark production via a chromo-magnetic moment at the LHC}

\author{Alexander Belyaev}
\affiliation{School of Physics \& Astronomy, University of Southampton, Southampton SO17 1BJ, UK}
\affiliation{Particle Physics Department, Rutherford Appleton Laboratory, Chilton, Didcot, Oxon OX11 0QX, UK}
%\email{a.belyaev@soton.ac.uk}

\author{R. Sekhar Chivukula}
\affiliation{UC San Diego, 9500 Gilman Drive,  La Jolla, CA 92023-0001, USA}
%\email{sekhar@ucsd.edu}

\author{Benjamin~Fuks}
\affiliation{Sorbonne Universit\'e, CNRS, Laboratoire de Physique Th\'eorique et Hautes \'Energies, LPTHE, F-75005 Paris, France}
\affiliation{Institut Universitaire de France, 103 boulevard Saint-Michel, 75005 Paris, France}
%\email{fuks@lpthe.jussieu.fr}

\author{Elizabeth H. Simmons}
\affiliation{UC San Diego, 9500 Gilman Drive,  La Jolla, CA 92023-0001, USA}
%\email{ehsimmons@ucsd.edu}

\author{Xing Wang}
\affiliation{UC San Diego, 9500 Gilman Drive,  La Jolla, CA 92023-0001, USA}
%\email{xiw006@ucsd.edu}

%%%%%%%%%%%%%%%%%%%%%%%%%%%%%%%%%%%%%%%%%%%%%%%%%%%%%%%%%%%%%%%%%%%%%%%%%%%
\begin{abstract}

Theories which provide a dynamical explanation for the large top-quark mass often include TeV-scale vector-like top-quark and bottom-quark partner states which can be potentially discovered at the LHC. These states are currently probed through model-independent searches for pair-production via gluon fusion, as well as through model-dependent complementary electroweak single production.
In this paper we study the potential to extend those searches for the partners of the third-generation Standard Model quarks on the basis of their expected chromomagnetic interactions.  We discuss how current searches for ``excited" bottom-quarks produced via $b$-gluon fusion through chromomagnetic interactions are relevant, and provide significant constraints. We then explore the region of the parameter space in which the bottom-quark partner is heavier than the top-quark partner, in which case the top-partner can be primarily produced via  the decay of the bottom-partner. Next, we probe the potential of the production of a single top-quark partner in association with an ordinary top-quark by gluon-fusion. Kinematically these two new processes are similar, and they yield the production of a heavy top partner and a lighter Standard Model state, a pattern which allows for the rejection of the associated dominant Standard Model backgrounds. We examine the sensitivity of these modes in the case where the top-partner subsequently decays to a Higgs boson and an ordinary top-quark, and we demonstrate that these new channels have the potential of extending and complementing the conventional strategies at LHC run III and at the high-luminosity phase of the LHC. In this last case, we find that partner masses that range up to about 3~TeV can be reached. This substantially expands the expected mass reach for these new states, including regions of parameter space that are inaccessible by traditional searches.

\end{abstract}

%\hspace*{100mm}{\large \tt PREPRINT NUMBER} \\

\maketitle
\flushbottom

%%%%%%%%%%%%%%%%%%%%%%%%%%%%%%%%%%%%%%%%%%%%%%%%%%%%%%
%%%%%%%%%%%%%%%%%%%%%%%%%%%%%%%%%%%%%%%%%%%%%%%%%%%%%%
\label{sec:intro}

%%%%%%%%% MASTE

\section{Introduction}

Theories which provide a dynamical explanation for the large top-quark mass often include TeV-scale vector-like top-quark ($T$) and bottom-quark ($B$) partner states which can be discovered at the LHC~\cite{Kaplan:1983fs,Kaplan:1991dc,Agashe:2004rs,Barnard:2013zea,Ferretti:2013kya} (see~\cite{Erdmenger:2020lvq, Erdmenger:2020flu} for recent works). These partner states can lie in various electroweak representations, and generally carry color charge. While $T\bar{T}$ or $B\bar{B}$ pair production via the strong interactions would be the dominant discovery mode for lighter states~\cite{Campbell:2009gj, Fuks:2016ftf}, single vector-like quark production offers another potentially relevant channel when the new quarks are heavy~\cite{Yue:2009cq,DeSimone:2012fs,Matsedonskyi:2014mna,Beauceron:2014ila,Liu:2015kmo,Backovic:2015bca,Liu:2017sdg,Kim:2018mks,Zidi:2018avr,Cacciapaglia:2018qep,Deandrea:2021vje}. Correspondingly, current searches at the LHC mainly focus on the QCD-induced production of a pair of vector-like partners~\cite{Aaboud:2017zfn,Aaboud:2017qpr,Aaboud:2018xuw,Aaboud:2018saj,Aaboud:2018xpj,Aaboud:2018wxv,Aaboud:2018pii,Sirunyan:2017pks,Sirunyan:2018qau,Sirunyan:2018omb,Sirunyan:2019sza,Sirunyan:2018yun,Sirunyan:2020qvb}. However, single production modes have also been recently considered for cases where the vector-like partner decays into a third generation quark and an electroweak scalar or vector boson~\cite{Aaboud:2018saj,Aaboud:2018ifs,Sirunyan:2017ynj,Sirunyan:2018fjh,Sirunyan:2018ncp,Sirunyan:2019xeh}. The associated limits lead to lower bounds on the vector-like quark masses of about 1--1.5~TeV, the exact bounds depend on the assumed vector-like quark decay pattern. In this paper we are largely concerned with this complementary single production process, but this time in the framework of scenarios that simultaneously feature both bottom-like and top-like partners.

In the case of the bottom partners, the most stringent bounds from single-production presently come from searches dedicated to signatures of ``excited" bottom quarks produced through $bg$-fusion via a chromomagnetic moment mixing operator of the form
\begin{equation}
\frac{g_s }{\Lambda}\  \bar{b}_{R,L} \sigma^{\mu\nu} G_{\mu\nu}  B_{L,R} \, ,
\label{eq:bGchromomagneticmoment}
\end{equation}
where $\Lambda$ parameterizes the compositeness scale associated with the partners. Such an interaction, which has usually been ignored, occurs naturally if ordinary quarks and their partners are both composite states arising from a dynamical theory of flavor and electroweak symmetry breaking and if they share the same constituents~\cite{Eichten:1983hw}. Current experimental limits targeting the single production of such an excited bottom quark that then decays into a $tW$ or $gb$ final state impose that this state must be heavier than about 1.2--1.6~TeV~\cite{CMS:2015rkn,CMS:2015jvf}. These limits, that can be applied to vector-like-quark setups in which the above operator is included, must however be modified when both bottom and top partners are present and form a single electroweak multiplet.

In this paper we explore such a new physics setup and study, in a twofold way, the potential to extend searches for top and bottom vector-like partners when the effects of the usually-neglected chromomagnetic interactions are considered. First, we study the region of the parameter space in which the bottom-quark partner is heavier than the top-quark partner. In this case the $B$-quark is produced via quark-gluon ($bg$) fusion through the interaction described above, and then decays to a $TW$ system. Second, as the $T$-quark is a color-triplet composite fermion, dimension-five gluon-$tT$ couplings analogous to the above bottom-quark chromomagnetic operator will be generated too, which reads as
\begin{equation}
\frac{g_s }{\Lambda}\  \bar{t}_{R,L} \sigma^{\mu\nu} G_{\mu\nu}  T_{L,R} \, .
\label{eq:chromo-single-T}
\end{equation}
This therefore provides an additional gluon-fusion mechanism for associated $Tt$ production.

Kinematically the $gb\to B \to TW$ and $gg\to tT$ processes are similar, as they yield the production of a heavy $T$-quark in association with a light state that could be either a $W$-boson or a Standard Model top quark. They therefore provide a distinctive signature. For example, we can consider the process $gg \to Tt$, where the top quark is dominantly produced at threshold.  The heavy $T$-quark can then for instance decay to a $th$ system, yielding a $tth$ final state. Both the top quark and the Higgs boson produced in the $T$-quark decay are in general highly boosted, and therefore appear in detectors as fat jets. While the bulk of the events contain two final-state top quarks, the very different kinematics of the two top quarks allows one to unambiguously distinguish them from one another. It becomes thus possible to combine the fat top jet and the fat Higgs jet (assuming an $h\to b\bar b$ decay) to seek evidence for an invariant mass peak corresponding to the vector-like $T$-quark, without worrying about any combinatorial issues originating from the presence of two final-state top quarks.

This specific signature also allows us, by its peculiar kinematic properties, to unravel the heavy quark signal from the associated key backgrounds. For instance, the production of a top-antitop pair together with an extra hard radiation leads to the same final state (two top quarks and a jet) and may have a large cross section, although only radiation of an unusually hard gluon would enable one of the top quarks to be as near-threshold as in the signal case. This background however has very different properties than the signal events. Combining the gluon and the softer top quark should reveal an invariant mass peak at the top mass $m_t$, while combining the gluon with the harder top quark should not exhibit any distinctive feature. On the other hand, background from $ttW$ production would be electroweak in rate and its kinematics would differ from those of the signal; there would be no reason here for one of the tops to be significantly boosted relative to the other.

In this work we show that for scales $\Lambda$ lying in the TeV-regime, the two considered signals complement conventional searches for vector-like-quark pair and single production, as well as those for single excited $b$-quark production. We demonstrate that the analysis of these production modes extend the reach of the LHC to vector-like quarks at run III and after the LHC high-luminosity (HL-LHC) phase in interesting regions of model parameter space. In practice, we focus on an illustrative scenario in which the top-partner subsequently decays to a Higgs boson and an ordinary top quark, although other decay channels would induce a final state with similar kinematic features. We find, for the HL-LHC, that top-partner masses ranging up to about 3~TeV could be reached, which substantially improves the LHC expectations for the considered new states, including in regions of parameter space that are inaccessible by traditional searches.

In the next section we introduce a simplified model of top- and bottom-partner states, delineate the parameter space of interest, and calculate relevant decay rates and production cross sections. In the third section we describe novel search strategies dedicated to chromomagnetic operators mixing vector-like partners and the Standard Model quarks, and list the corresponding dominant backgrounds. We then explore the associated discovery reach at the LHC and its high-luminosity operation phase. Conclusions and directions for future work are given in the fourth section.

%%%%%%%%%%%%%%%%%%%%%%%%%%%%%%%%%%%%%%%%%%%%%%%%%%%%%%

\label{sec:model}

\section{Model}
In this paper we illustrate the potential of chromomagnetic moment interactions for extending collider searches for third-generation partner quarks by focusing on the simplest composite Higgs effective theory for the third-generation-quark sector that incorporates partial compositeness. In order to define our theoretical framework, we rely on a setup similar to that of Ref.~\cite{Vignaroli:2012nf}. In section~\ref{sec:fields}, we describe the fermionic field content of the model and discuss the effective dimension-five chromomagnetic interactions in which composite fermions are generically involved. In section~\ref{sec:spectra}, we introduce the model parameter space and investigate the mass spectrum of the new states, as well as predictions for the associated decay patterns.

%%%%%%%%%%%%%%%%%%%%%%%%%%%%%%%%%%%%%%%%%%
\subsection{Field Content and Fermion Eigenstates}\label{sec:fields}
We start from the Standard Model (SM) field content that we complement with a small set of new composite (electroweak) doublets and singlets of vector-like fields directly associated with the generation of the top-quark mass,
\begin{equation}
Q^0_{L,R}=
\begin{pmatrix}
	T^0_{L,R} \\ B^0_{L,R}
\end{pmatrix}
\ \ \text{and} \ \ \tilde{T}^0_{L,R}\, .
\end{equation}
In this notation, the superscript ``0'' indicates that the fields are gauge eigenstates, in constrast with the mass eigenstates $T_1$, $T_2$ and $B_1$ that are defined below. These new gauge eigenstates couple to their elementary Standard Model top-quark and bottom-quark counterparts,
\begin{equation}
q_L = \begin{pmatrix}
t^0_L \\ b^0_L
 \end{pmatrix}
 \ \ \text{and}  \ \ t^0_R\, ,
\end{equation}
via mass mixings as described by the Lagrangian\footnote{We consider the bottom-quark mass to be zero, and neglect the sector of the model associated with the bottom-quark mass generation. The bottom quark mass can be generated via a similar mechanism as for the top-quark by adding a second composite bottom quark, which is however heavy and does not affect the phenomenology discussed here.}
\begin{equation}
{\cal L}_{\rm mass} =
  -M_Q \overline{Q^0_L} Q^0_R
  -M_{\tilde{T}} \overline{\widetilde{T}^0_L} \widetilde{T}^0_R
  - \Big(
      y^* (\overline{Q^0_L} \cdot \Phi^\dag) \widetilde{T^0_R}
    + \Delta_L \overline{q^0_L} Q^0_R 
    + \Delta_R \overline{t^0_R} \widetilde{T}^0_L
    + {\rm H.c.}
  \Big)\, .
\label{eq:mixing} \end{equation}
In this expression, the field $\Phi$ stands for the Standard Model Higgs doublet (that has a composite origin in our framework), the dot product refers to the invariant product of two fields lying in the (anti)fundamental representation of $SU(2)_L$, and the $M_Q$, $M_{\tilde T}$, $\Delta_L$ and $\Delta_R$ quantities denote the strength of the various bilinear mass mixing parameters. The Yukawa coupling $y^*$ describes the mixing of the doublet and singlet vector-like quarks via their interactions with the Higgs doublet. As we assume that the Higgs field has a composite origin, we neglect interactions that allows the Higgs to directly couple to the elementary quarks (like $\bar{Q}_L\phi^ct_R$ for instance). The mass terms from Eq.~\eqref{eq:mixing} can be more conveniently written in a matrix form as
\begin{equation}
	{\cal L}_{\rm mass} = 
	-\begin{pmatrix}
		\overline{t^0_L} & \overline{T^0_L} & \overline{\tilde{T}^0_L}
	\end{pmatrix}
	{\cal M}_t
	\begin{pmatrix}
		t^0_R \\ T^0_R \\ \widetilde{T}^0_R
	\end{pmatrix}
  -
\begin{pmatrix}
	\overline{b^0_L} & \overline{B^0_L}
\end{pmatrix}
{\cal M}_b
\begin{pmatrix}
	b^0_R \\
	B^0_R
\end{pmatrix} \, ,
\end{equation}
where ${\cal M}_t$ and ${\cal M}_b$ are the fermion mass matrices defined by
\begin{equation}
	{\cal M}_t = 
	\begin{pmatrix}
		0 & \Delta_L & 0\\
		0 & M_Q & m\\
		\Delta_R & m & M_{\widetilde{T}}
	\end{pmatrix}
	\qquad\text{and} 
\qquad
{\cal M}_b = 
\begin{pmatrix}
	0 & \Delta_L\\
	0 & M_Q
\end{pmatrix}
\label{eq:mass-matrix}\ \ .
\end{equation}
The element $m$ appearing in the ${\cal M}_t$ matrix arises from the Yukawa interaction after electroweak symmetry breaking. It is given by
\begin{equation}
   m=\frac{y^* v}{\sqrt{2}}~,
\end{equation}
where $v \approx 246$ GeV is the vacuum expectation value of the SM Higgs field. The fermion physical masses and the corresponding eigenvectors are derived by diagonalizing the squared mass matrices ${\cal M} {\cal M}^\dagger$ for the left-handed eigenvectors, and correspondingly the squared mass matrices ${\cal M}^\dagger {\cal M}$ for the right-handed ones. In general, this diagonalization can only be done numerically. The left-handed and right-handed components of the mass eigenstates $t$, $T_1$, $T_2$, $b$ and $B_1$ are generically given by
\begin{equation}
	\begin{pmatrix}
		t_L \\ T_{1L} \\ T_{2L}
	\end{pmatrix}
	= O^t_{L}
	\begin{pmatrix}
		t^0_L \\ T^0_L \\ \tilde{T}^0_L
	\end{pmatrix},\ \
	\begin{pmatrix}
		t_R \\ T_{1R} \\ T_{2R}
	\end{pmatrix}
	= O^t_{R}
	\begin{pmatrix}
		t^0_R \\ T^0_R \\ \tilde{T}^0_R
	\end{pmatrix}, \ \
	\begin{pmatrix}
		b_L \\ B_{1L}
	\end{pmatrix}
	= O^b_{L}
	\begin{pmatrix}
		b^0_L \\ B^0_L
	\end{pmatrix},\ \
	\begin{pmatrix}
		b_R \\ B_{1R}
	\end{pmatrix}
	=
	\begin{pmatrix}
		b^0_R \\ B^0_R
	\end{pmatrix}\, ,
\label{eq:mixingbis}\end{equation}
after introducing the mixing matrices $O^t_L$, $O^t_R$ and $O^b_L$.

Partial compositeness generally predicts the generation of dimension-five chromomagnetic interactions at the electroweak scale,
\be
{\cal L}_{\rm chromo}=\frac{g_s }{\Lambda}\  \overline{{\cal Q}_L} \sigma^{\mu\nu} G_{\mu\nu} {\cal Q}_R + {\rm H.c.}\,  ,
\label{eq:chromomag} \ee
where $\sigma_{\mu\nu} = i(\gamma_\mu\gamma_\nu-\gamma_\nu\gamma_\mu)/2$, and ${\cal Q}_L$ and ${\cal Q}_R$ denote any of the considered left-handed and right-handed new physics gauge eigenstates (${\cal Q} = Q^0$, $\tilde T^0$). The gluon field strength tensor reads $G_{\mu\nu}=G^A_{\mu\nu} T_A$, where the matrices $T_A$ are the fundamental representation matrices of $SU(3)$ (all considered states being color triplets), and we have assumed that the compositeness scale $\Lambda$ is the same for all considered vector-like quarks, which is a natural simplifying assumption. As a result of the mixing of the SM quarks with their composite  partners given by Eq.~\eqref{eq:mixingbis}, the Lagrangian (\ref{eq:chromomag}) gives rise to the  ``off-diagonal" chromomagnetic interactions involving a single third-generation SM quark and a single vector-like quark. These are given by
\be\bsp
  {\cal L}_{t} =&\ \frac{g_s}{\Lambda} G_{\mu\nu}\  \bigg[
    ({O^t_L}_{12}{O^t_R}_{22}+{O^t_L}_{13}{O^t_R}_{23})  \overline{T}_{1R} \sigma^{\mu\nu} t_L +
    ({O^t_R}_{12}{O^t_L}_{22}+{O^t_R}_{13}{O^t_L}_{23}) \overline{T}_{1L} \sigma^{\mu\nu} t_R \\
  & \quad +
    ({O^t_L}_{13}{O^t_R}_{33}+{O^t_L}_{12}{O^t_R}_{32}) \overline{T}_{2R} \sigma^{\mu\nu} t_L +
    ({O^t_H}_{13}{O^t_L}_{33}+{O^t_R}_{12}{O^t_L}_{32}) \overline{T}_{2L} \sigma^{\mu\nu} t_R
  \bigg] + {\rm H.c.} \, , \\
  {\cal L}_{b} =&\ \frac{g_s}{\Lambda} G_{\mu\nu}\  \bigg[
    {O^b_L}_{12} \overline{B}_{1R} \sigma^{\mu\nu} b_L
  \bigg] + {\rm H.c.} \, ,
\esp\label{eq:bbg}\ee
where all mixing matrices are taken real for simplicity. These off-diagonal chromomagnetic interactions open new opportunities for the exploration of vector-like quark physics at colliders, beyond those already accessible and investigated today. The advantage of single production in comparison to the QCD pair production mechanism is the smaller phase space suppression; the advantage of the chromomagnetic-moment-induced single production process relative to electroweak single production is its enhancement due to the strong coupling and a gluon density in the initial state. The chromomagnetic operator in Eq.~(\ref{eq:chromomag}) also modifies the ``diagonal'' $gt\bar{t}$ and $gT_1\bar{T}_1$  QCD interactions, and can thus be probed through $t\bar{t}$ and $T_1\bar{T}_1$ pair production~\cite{BuarqueFranzosi:2019dwg}.

In principle, scenarios such as those investigated in this work are constrained by electroweak precision tests. However, we adopt a bottom-up approach in which we consider a simplified (and therefore not UV-complete) model. To accommodate the electroweak constraints, one could extend our setup, for example, by incorporating a realization of the custodial symmetry. While such extensions would result in richer mass spectra, our model still captures the important phenomenology of the top-partner and bottom-partner sectors which we address in this work. In particular, the $Zb\bar{b}$ coupling (which is usually the source of stringent electroweak constraints) is not modified as both elementary and composite bottom quarks lie in an $SU(2)_L$ doublet.

\subsection{Mass Spectra and Vector-Like Quark Decays}\label{sec:spectra}
Once we account for the fact that the mass of the SM top quark is known, the theoretical setup defined in section~\ref{sec:fields} is described by five independent free parameters, that we choose to be
\begin{equation}
\bigg\{ \epsilon_L=\frac{\Delta_L}{M_Q}, \ \ \epsilon_R=\frac{\Delta_R}{M_{\tilde{T}}}, \ \  m_{T_1}, \ \  \frac{M_Q}{M_{\tilde{T}}} \ \ ,  \ \ \Lambda\bigg\}\, . \label{eq:epsilons}
\end{equation}
The Yukawa coupling $y^*$ that appears in Eq.~(\ref{eq:mixing}), as well as the  physical masses of the composite quarks $m_{T_2}$ and $m_{B_1}$, are then fixed by the SM top quark mass measurements and by the parameters of the model in Eq.~(\ref{eq:epsilons}). We investigate scenarios in which the doublet-to-singlet mass ratio $M_Q/M_{\tilde{T}}$ takes discrete values equal to 1 or 2. This parameter controls both the decoupling of the heavier $T_2$ state with respect to the $T_1$ and $B_1$ states , and the singlet/doublet nature of the $T_1$ and $T_2$ quarks. For the parameters we choose in this work, the hierarchy $m_{T_1}<m_{B_1}<m_{T_2}$ is always realized\footnote{There exist scenarios where $m_{B_1}<m_{T_1}<m_{T_2}$ if one chooses $M_{\tilde{T}}>M_Q$ with a large $\epsilon_R$ value. We have numerically checked that the mass splitting in such scenarios is small, $m_{T_1}-m_{B_1}\sim\mathcal{O}(10~{\rm GeV})$, and does not result in appreciable decay rate of $T_1\rightarrow B_1 W$. Therefore, we ignore such scenarios since they do not lead to any qualitatively different collider signature.}. For $M_Q/M_{\tilde{T}}=1$, all mass splittings featured by the physical vector-like states solely arise from the off-diagonal $\epsilon$ and $m$ terms of the mass matrices, whilst for $M_Q/M_{\tilde{T}}=2$ an extra source of splitting originates from the large difference between the diagonal elements of the ${\cal M}_t$ and ${\cal M}_b$ matrices. Those two choices are sufficient to discuss the distinct search strategies associated with the chromomagnetic operators of Eq.~\eqref{eq:bbg} and that we propose in the present work. We therefore end up with a four-dimensional continuous parameter space corresponding to each choice of $M_Q/M_{\tilde{T}}$.

\begin{figure}[t]
\centering
\begin{subfigure}{0.45\textwidth}
	\includegraphics[width=\textwidth]{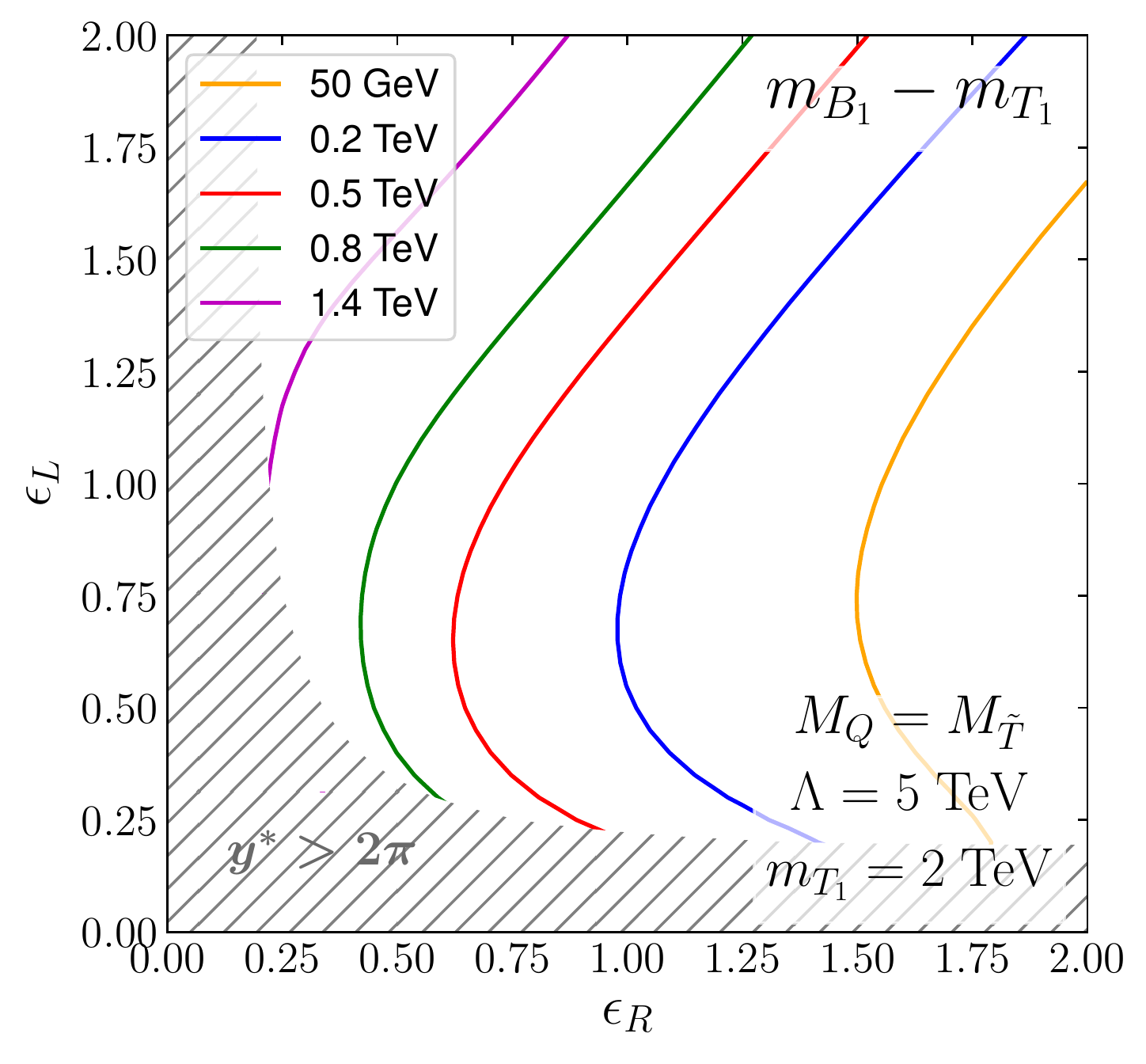}%	
	\caption{}
\end{subfigure}
\begin{subfigure}{0.45\textwidth}
	\includegraphics[width=\textwidth]{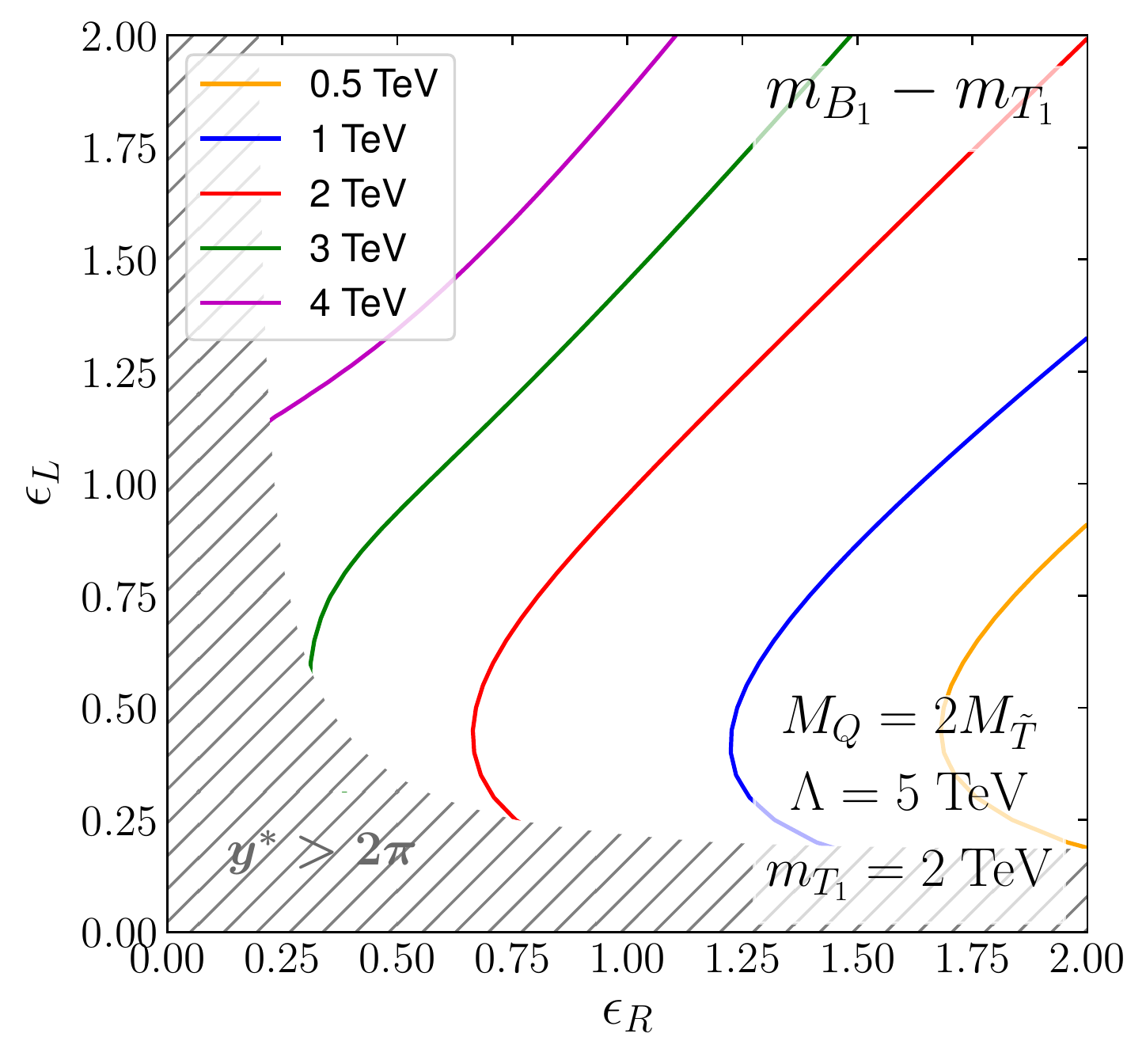}	
	\caption{}
\end{subfigure}
\caption{Illustrative mass spectrum of the lighter vector-like partners $T_1$ and $B_1$. We present the mass difference $m_{B_1}-m_{T_1}$ in the $(\epsilon_L, \epsilon_R)$ plane for $\Lambda=5$~TeV, $m_{T_1}=2$~TeV and (a)  $M_Q/M_{\tilde{T}}=1$ and (b) $M_Q/M_{\tilde{T}}=2$. The hashed regions are excluded by requiring $y^* < 2\pi$ to comply with perturbativity constraints.}
\label{figs:mass} \end{figure}
Illustrative mass spectra are shown in Figure~\ref{figs:mass}, in which we present iso-contours for the $m_{B_1}-m_{T_1}$ mass difference in the $(\epsilon_L, \epsilon_R)$ plane for $\Lambda=5$~TeV, $m_{T_1}=2$~TeV and $M_Q/M_{\tilde{T}}=1$ $(2)$ in panels (a) and (b), respectively, of the figure. We observe that the $m_{B_1}-m_{T_1}$ mass splitting is anti-correlated with $\epsilon_R$, whereas it is on the contrary correlated with $\epsilon_L$. The two vector-like states $B_1$ and $T_1$ hence become quasi-degenerate for large $\epsilon_R$ values, as long as $\epsilon_L$ is small enough. For instance, the $m_{B_1}-m_{T_1}$ mass difference drops down to a few dozen (hundred) GeV for $\epsilon_R \simeq 2$ if $\epsilon_L \simeq 0.6$ (0.5) for configurations featuring $M_Q/M_{\tilde{T}}=1$ (2). At such $\epsilon$ values, the mass splitting is moreover minimal. In contrast, very large splitting reaching 1~TeV or more take place for $\epsilon_L \simeq 2$ and small values of $\epsilon_R$. In the latter configuration, the $T_1$ state is mostly of a weak-singlet nature, while the doublet-like states $B_1$ and $T_2$ are both much heavier. Moreover, there is no strong dependence of the mass spectra on $M_Q/M_{\tilde{T}}$, although for greater $M_Q/M_{\tilde{T}}$ values, the variations of the $m_{B_1}-m_{T_1}$ mass splitting become milder.

\begin{figure}[t]
	\centering
\begin{subfigure}{0.45\textwidth}
	\includegraphics[width=\textwidth]{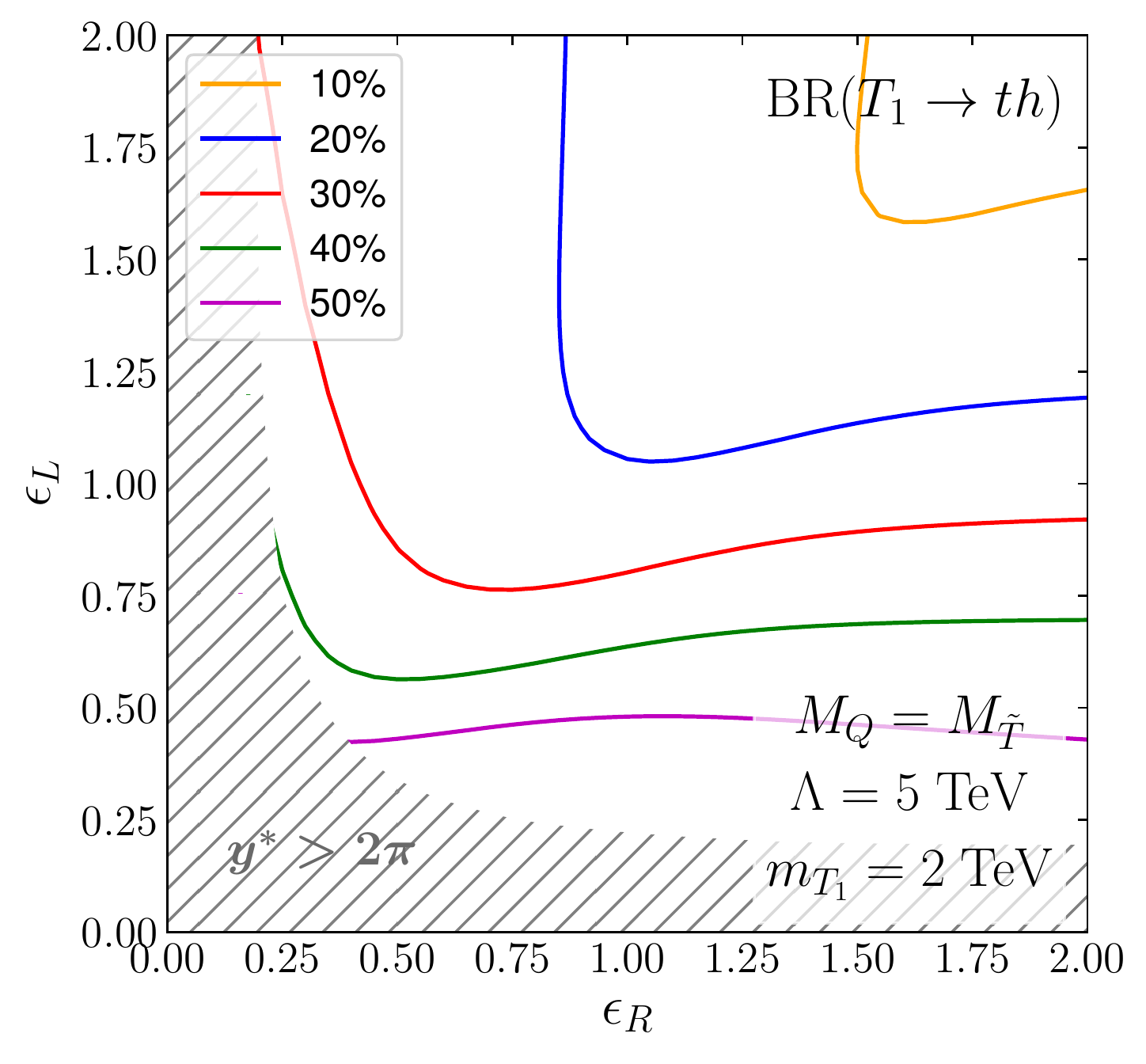}%	
	\caption{}
\end{subfigure}
\begin{subfigure}{0.45\textwidth}
	\includegraphics[width=\textwidth]{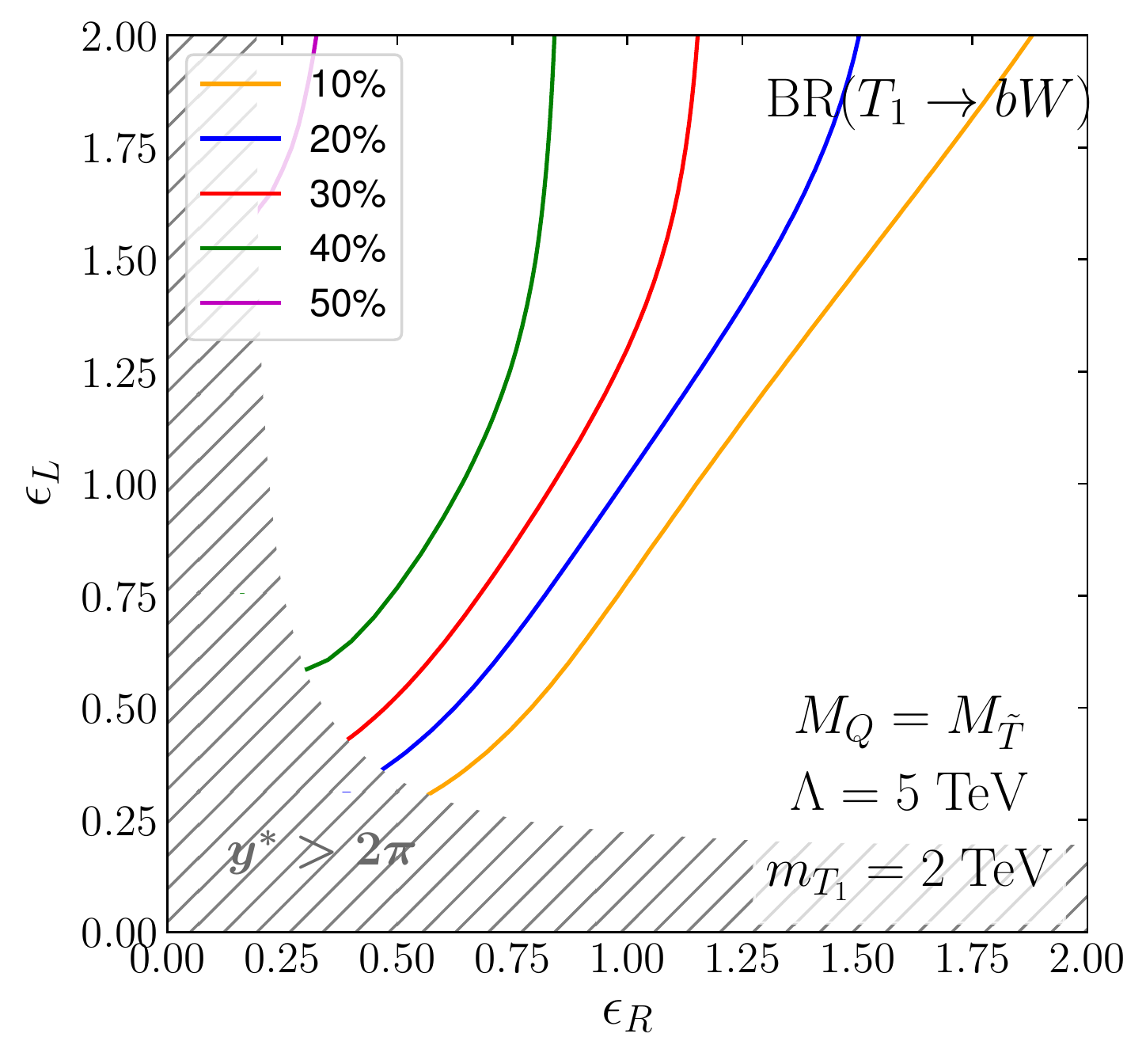}
	\caption{}
\end{subfigure}
\begin{subfigure}{0.45\textwidth}
	\includegraphics[width=\textwidth]{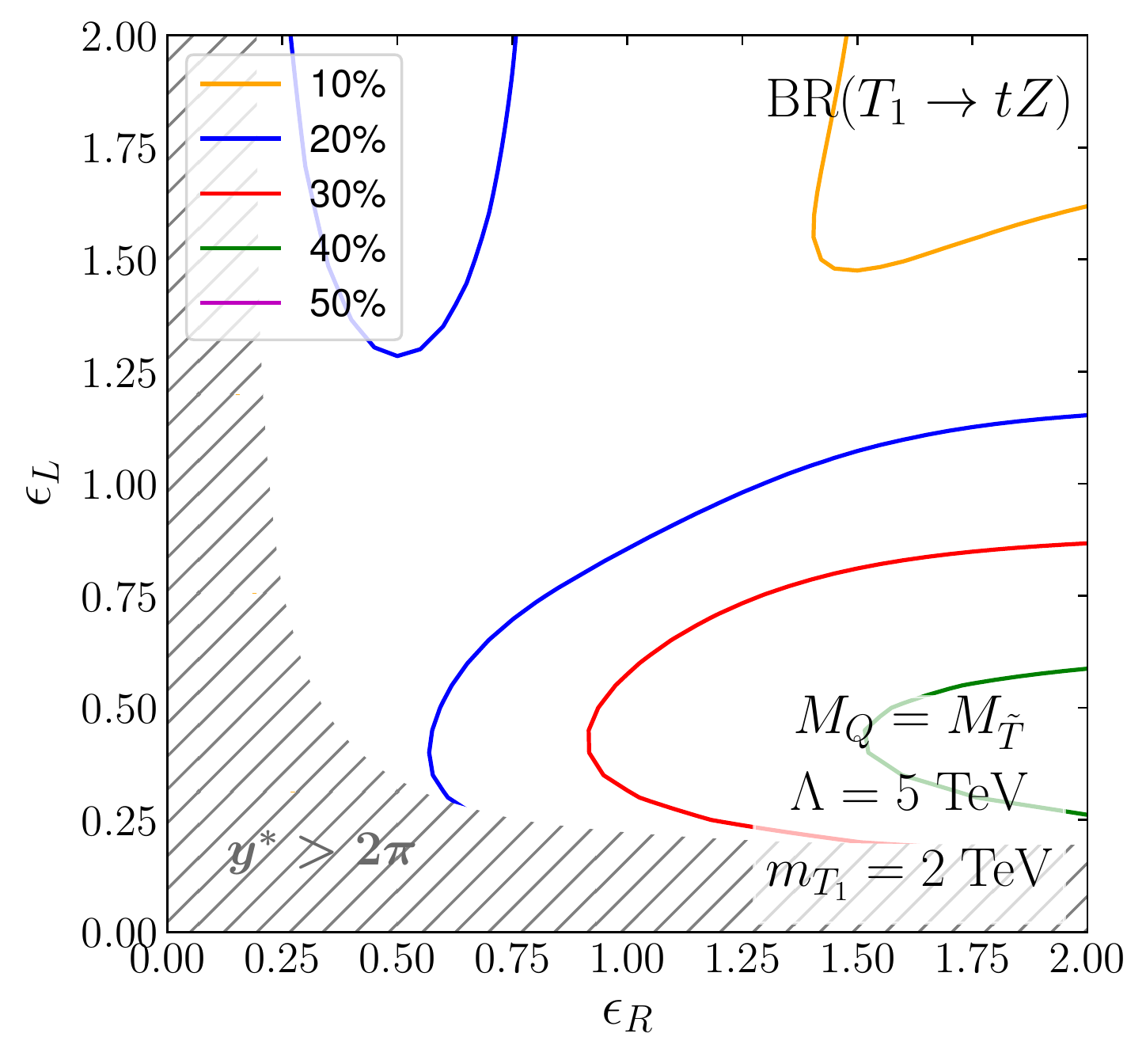}%	
	\caption{}
\end{subfigure}
\begin{subfigure}{0.45\textwidth}
	\includegraphics[width=\textwidth]{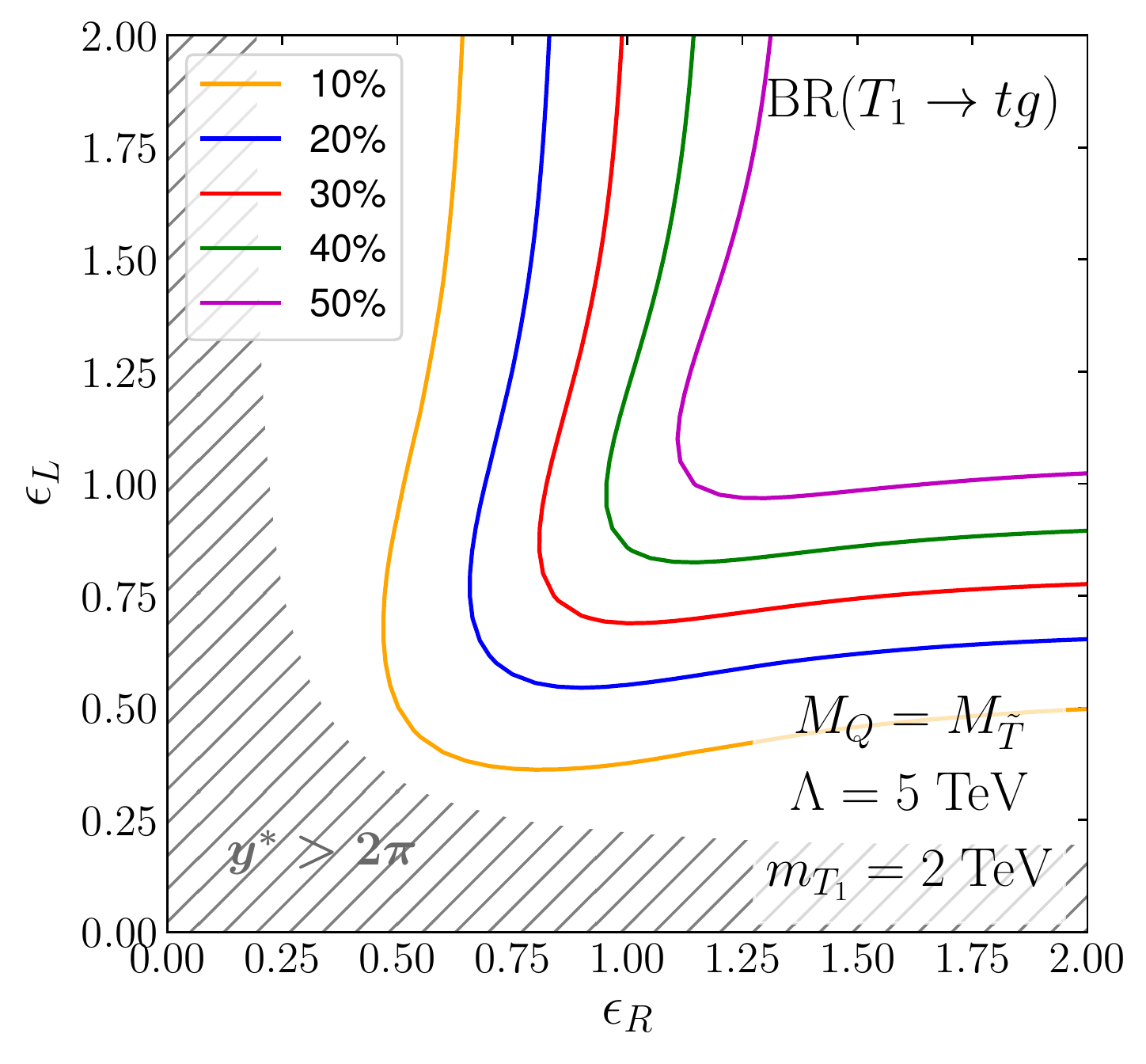}%	
	\caption{}
\end{subfigure}
\caption{Isocontours associated with the branching ratios of the $T_1$ quark when it decays to a $th$~(a), $bW$ (b), $tZ$ (c) and $tg$ (d) system. The results are presented in the $(\epsilon_L, \epsilon_R)$ plane, for $\Lambda=5$~ TeV, $m_{T_1}=2$~TeV and $M_Q/M_{\tilde{T}}=1$.}\label{figs:BrT}
\end{figure}
The mass spectrum is strongly correlated with branching ratios of the different states. In Figure~\ref{figs:BrT} we present contour levels for the branching ratios of the $T_1$ quark when it decays to a $th$ system (a), a $bW$ system (b), a $tZ$ system (c) and a $tg$ system (d), with the last decay mode occurring via the dimension-five chromomagnetic operators. The results are presented once again in the $(\epsilon_L, \epsilon_R)$ plane for a fixed value of the other parameters of the model, for which we adopt  $\Lambda=5$~TeV, $m_{T_1}=2$~TeV and $M_Q/M_{\tilde{T}}=1$.

The pattern of the branching ratios BR$(T_1 \to th)$ and BR$(T_1 \to tg)$ is found to be (almost) symmetric with respect to the $\epsilon_L=\epsilon_R$ diagonal line,  indicating that these branching ratios are proportional to the $\epsilon_L \epsilon_R$ product. This originates from the chiral structure of the corresponding interactions that both connect a left-handed fermion to a right-handed one. On the other hand, the branching ratio BR$(T_1 \to th)$ is proportional to the squared Yukawa coupling $y^{*2}$. However, such a coupling decreases with the increase of $\epsilon_L=\epsilon_R$ values along the diagonal, this behaviour stemming from the requirement that the lightest mass eigenvalue of the top-quark mass matrix ${\cal M}_t$ be equal to the mass of the SM top quark. As a consequence, the BR$(T_1 \to tg)$ branching ratio increases along this diagonal for increasing $\epsilon$ values, while the branching ratio BR$(T_1 \to th)$ decreases. Conversely, in regions where $y^*$ is large, namely near the boundary of the parameter space allowed by perturbativity (which is defined by enforcing that $y^*< 2\pi$), the BR$(T_1 \to th)$ branching ratio is dominant and can even exceed 50\%.

At the same time the BR$(T_1 \to bW)$ branching fraction increases along the direction of the orthogonal diagonal, {\it i.e.}, when evolving from the bottom-right to the top-left corner of the represented region of the $(\epsilon_L, \epsilon_R)$ plane. This branching ratio becomes larger because the strength of the coupling of the $T_1$ fermion to the $W$-boson and the SM bottom quark gets larger, as this strength is approximately proportional to the $\epsilon_L$ parameter. Subsequently, the figure illustrates the expected behaviour of BR$(T_1 \to bW)\sim\epsilon_L^2$.

Finally, the BR$(T_1 \to tZ)$ branching, that is proportional to the sum $\epsilon_L^2 + \epsilon_R^2$ and that is at the same time independent from their product $\epsilon_L \epsilon_R$, is only large when one of the mixing parameter is large and the other one is small. In such a configuration, all other decay channels are suppressed. However even in this case the BR$(T_1 \to tZ)$ branching ratio does not exceed the BR$(T_1 \to th)$ one. This therefore motivates us to choose the $T_1 \to th$ channel as a key decay mode for our exploration of the vector-like quark chromomagnetic moments in Section~\ref{sec:pheno}.

\begin{figure}[t]
	\begin{subfigure}{0.45\textwidth}
		\includegraphics[width=\textwidth]{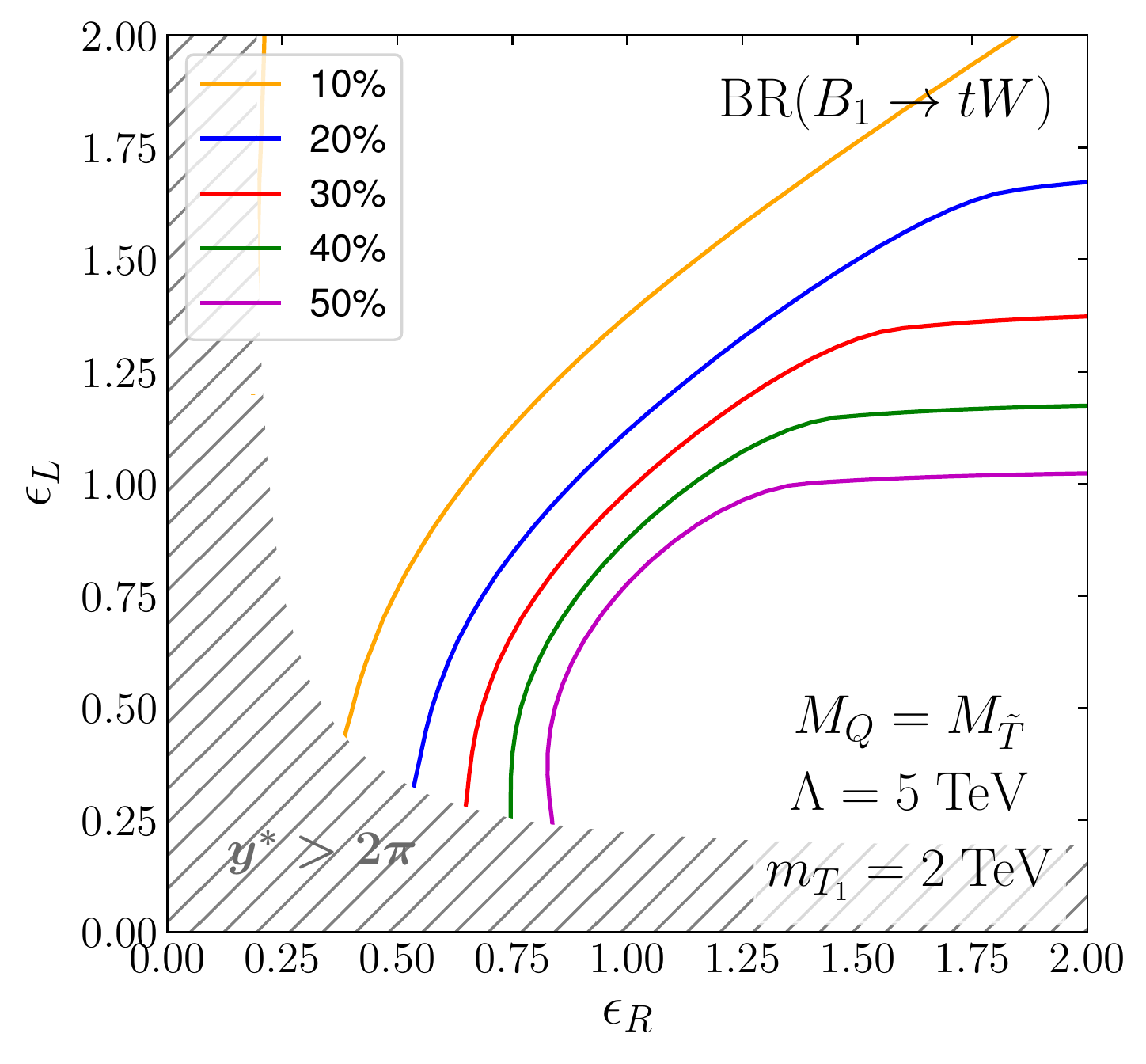}%	
		\caption{}
	\end{subfigure}
	\begin{subfigure}{0.45\textwidth}
		\includegraphics[width=\textwidth]{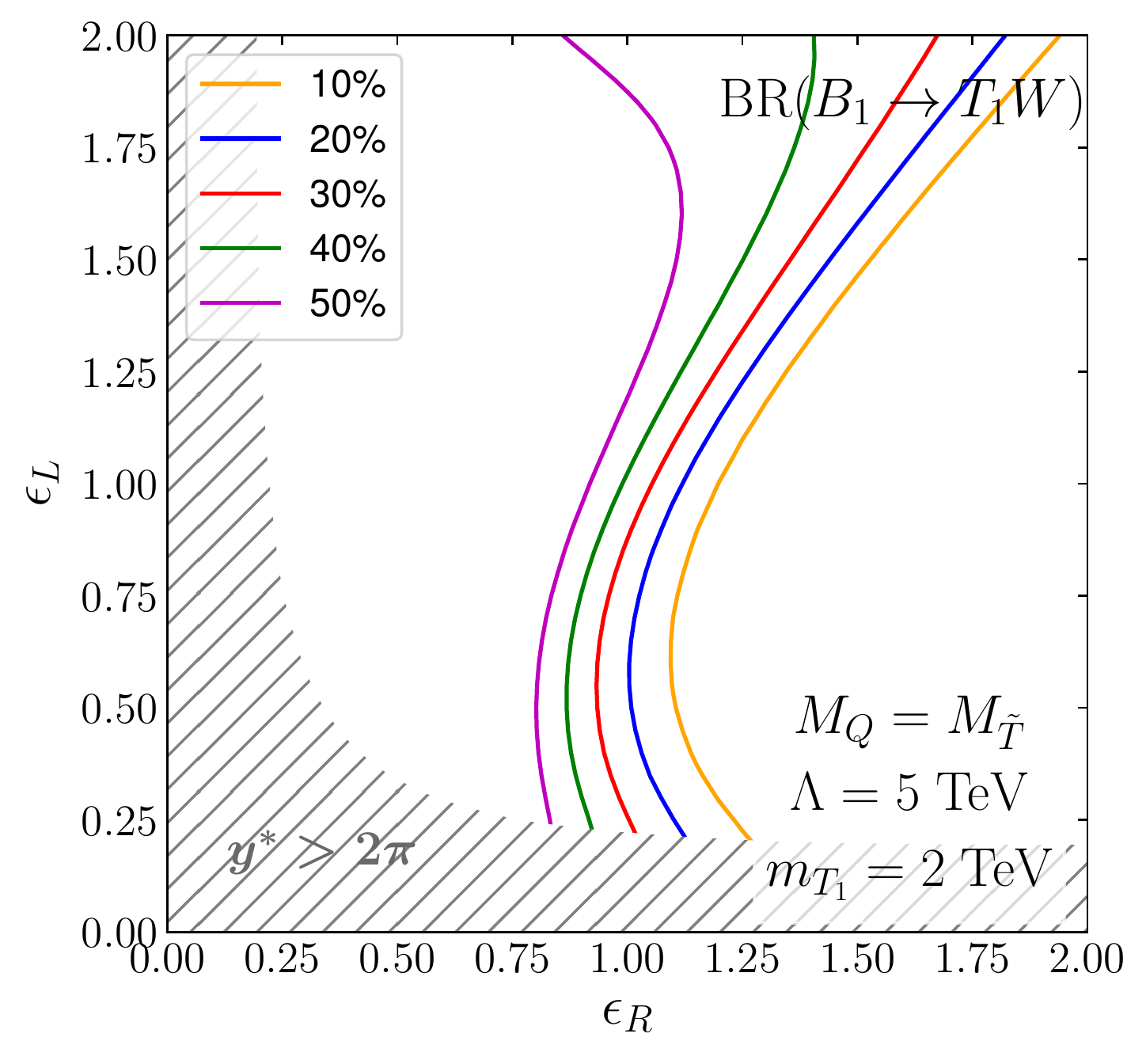}
		\caption{}
	\end{subfigure}
	\begin{subfigure}{0.45\textwidth}
		\includegraphics[width=\textwidth]{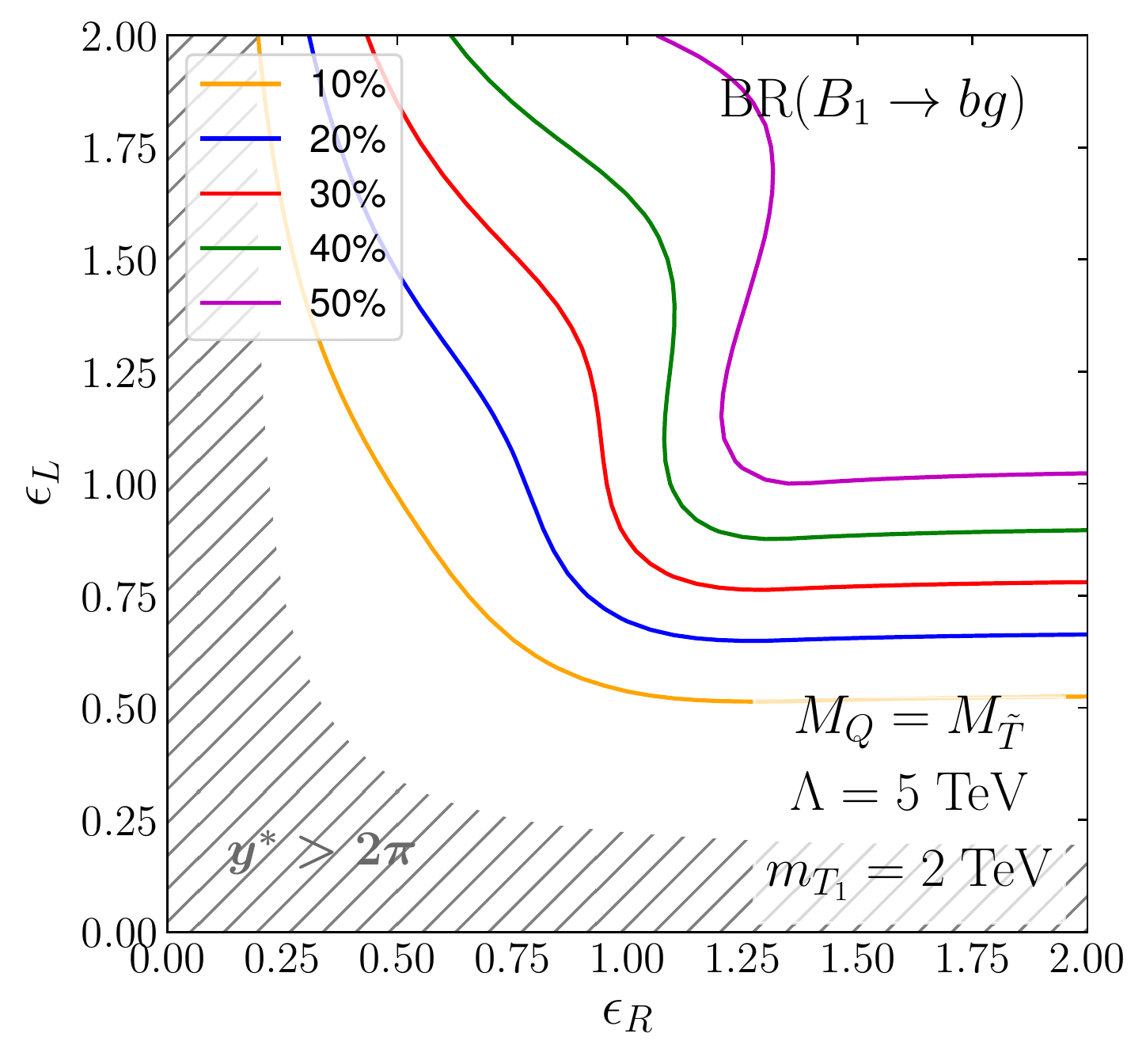}%	
		\caption{}
	\end{subfigure}
\caption{Isocontours associated with the branching ratios of the $B_1$ quark when it decays to a $tW$~(a), $T_1W$ (b) and $bg$ (c) system. The results are presented in the $(\epsilon_L, \epsilon_R)$ plane, for $\Lambda=5$~ TeV, $m_{T_1}=2$~TeV and $M_Q/M_{\tilde{T}}=1$.}\label{figs:BrB}
\end{figure}

We also study the decay pattern of the $B_1$ partner of the bottom quark as a function of the $\epsilon$ mixing parameters. The results are shown in Figure~\ref{figs:BrB} for the $B_1\to tW$ (a), $T_1W$ (b) and $bg$ (c) decays, and they are presented again in the $(\epsilon_L, \epsilon_R)$ plane for scenarios featuring  $\Lambda=5$~ TeV, $m_{T_1}=2$~TeV and $M_Q/M_{\tilde{T}}=1$.\footnote{The decays $B_1 \to bh$ and $B_1 \to bZ$ are absent in the limit of a vanishing $b$-quark mass.}

In the first two subfigures, we can observe the interplay of the BR$(B_1\to tW)$ and BR$(B_1\to T_1W)$ branching fractions. Whereas these two branching ratios are both proportional to the same gauge coupling, BR$(B_1\to tW)$ is suppressed by the mixing, while BR$(B_1\to T_1W)$ is not. This makes the $B_1\to T_1W$ decay dominant as soon as it has enough phase space. Consequently, the pattern in the $(\epsilon_L, \epsilon_R)$ plane featured by the $B_1\to T_1W$ branching ratio is correlated with the one of the $m_{B_1}-m_{T_1}$ mass splitting  discussed in Figure~\ref{figs:mass}. Because of the same reason the pattern of the BR$(B_1\to tW)$ branching fraction is anti-correlated with the $m_{B_1}-m_{T_1}$ mass splitting and reaches its maximum in the bottom-right corner of the $(\epsilon_L,\epsilon_R)$ parameter space, in which $m_{B_1}-m_{T_1}$ is minimal and the $B_1\to T_1W$ decay is either closed or strongly suppressed. The $B_1\to T_1 W$ decay therefore naturally dominates when the mixing is not too large and when it is kinematically favored, and the $B_1\to tW$ decay is only significant for a compressed spectrum and for a mixing combination exhibiting a large $\epsilon_R$ value and  a small $\epsilon_L$ value.

In the last subfigure we observe that $B_1\to bg$  decays exhibit a similar pattern to the one of the $T_1\to tg$ decay, the corresponding decay width being proportional to $\epsilon_L \epsilon_R$. Such a decay channel is thus only relevant for scenarios featuring two large mixing parameters.

In the present section, we have studied typical $B_1$ and $T_1$ mass splitting configurations and decay patterns. We have shown that the heavy top quark $T_1$ often decays into a system comprising a SM Higgs boson and a SM top quark, and that the heavier (but not always much heavier) $B_1$ quark has two potential dominant decay modes into a $tW$ and a $T_1W$ system. In the next section, we will study the phenomenology of the our model for these cases, investigating the dependence of the existing constraints on the mixing parameters and new discovery modes of the model stemming from non-vanishing dimension-five chromomagnetic operator as depicted by the Lagrangian~\eqref{eq:chromomag}.

%%%%%%%%%%%%%%%%%%%%%%%%%%%%%%%%%%%%%%%%%%%%%%%%%%%%

\section{Searches at the LHC}
\label{sec:pheno}

\newcommand{\figdirA}{figs/figs_B1_QT_1}
\newcommand{\figdirB}{figs/figs_B1_QT_2}

Top and bottom partners have been actively searched for at the LHC, and as discussed above the most stringent current constraints come both from QCD pair production and electroweak single production. Furthermore, a model-dependent search for excited $b$ quarks can be recast to get additional (and quite strong) bounds on the bottom-quark partners. In the present study, we examine the impact of the chromomagnetic operators in \eqref{eq:chromomag} and \eqref{eq:bbg}. The considered operators allow for the single production of the bottom and top partners with a coupling strength of order of the QCD coupling $g_s$, although the related processes undergo a suppression stemming from the higher-dimensional nature of the chromomagnetic moments. In this section, we focus on the associated LHC phenomenology and we investigate in particular the sensitivity of the future high-luminosity phase of the LHC (HL-LHC) to these interactions.

\subsection{Vector-Like Quark Single and Pair Production}\label{sec:pairsingle}

\begin{figure}
\centering
\begin{subfigure}{0.45\textwidth}
\includegraphics[width=\textwidth]{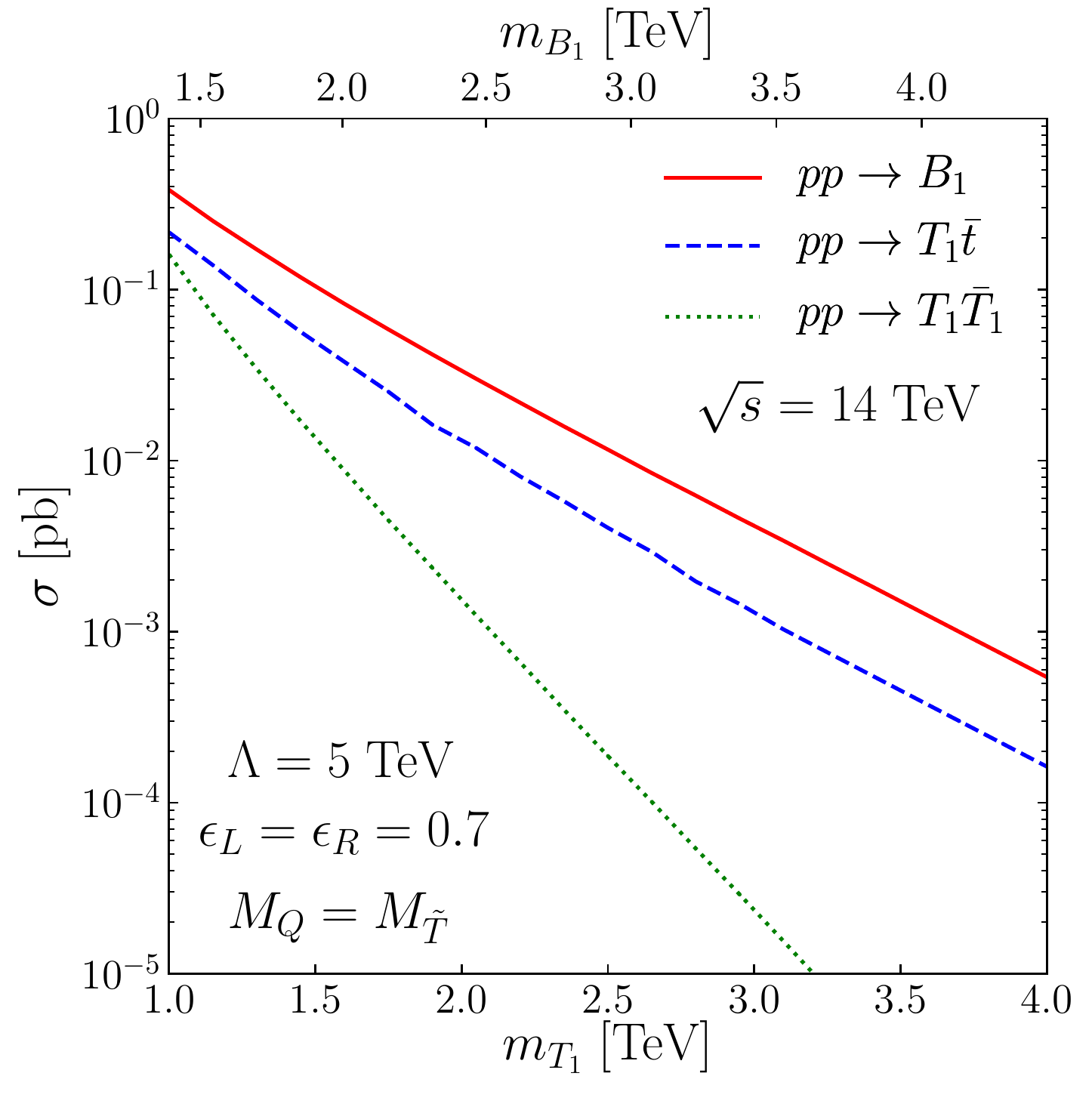}
\caption{}
\end{subfigure}
\begin{subfigure}{0.48\textwidth}
\includegraphics[width=\textwidth]{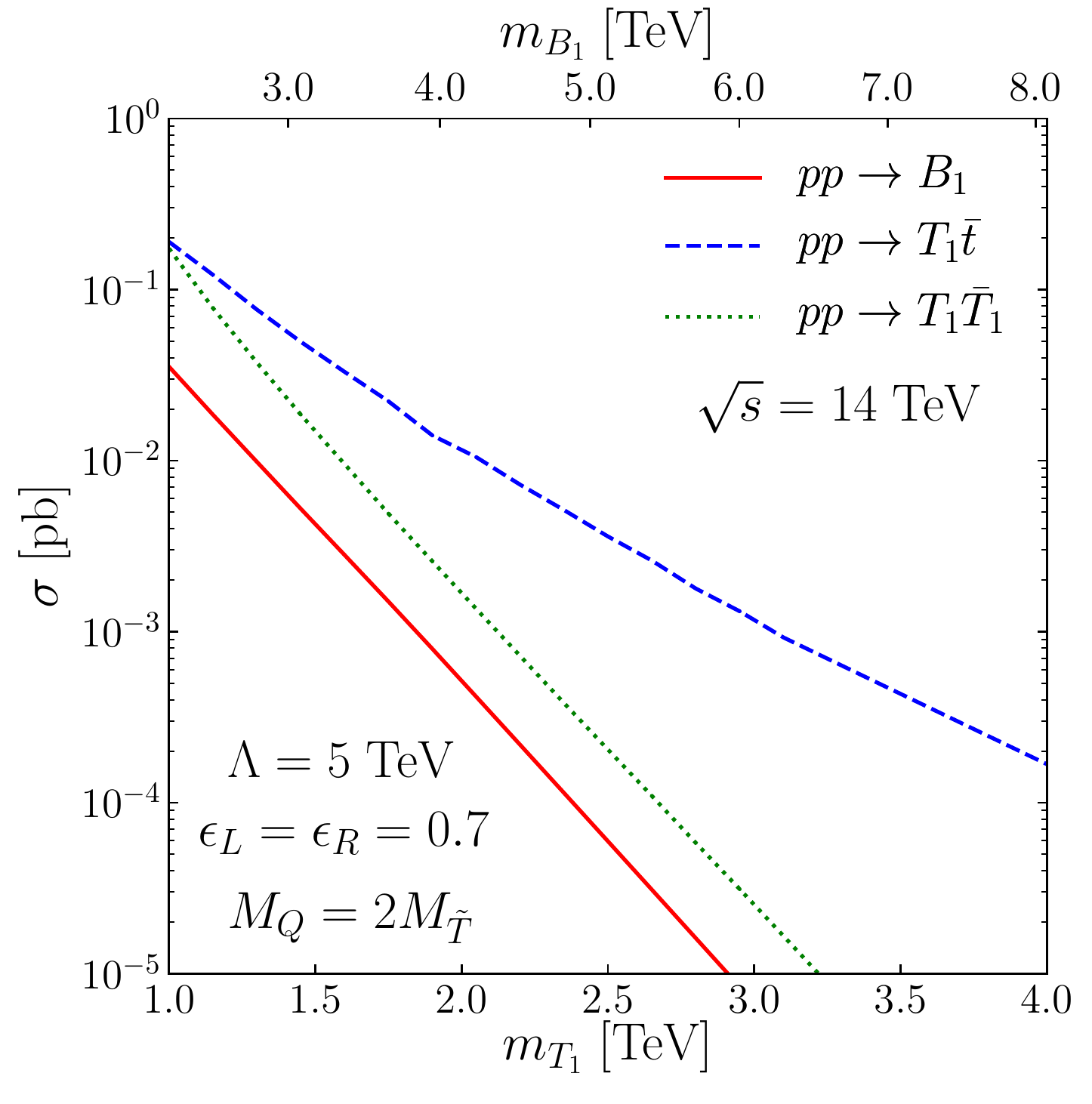}
\caption{}
\end{subfigure}
\caption{Production cross sections for the processes $pp\to B_1$ (solid red), $pp\to T_1\bar t$ (dashed blue) and $pp\to T_1\bar T_1$ (dotted green) at $\sqrt{s}=14$~TeV. The first two processes occur through the chromomagnetic moment operators, whereas the last process is a pure QCD process. We consider as a benchmark scenario a configuration in which $\epsilon_L = \epsilon_R = 0.7$, $\Lambda = 5$ TeV, and in which (a)  $M_Q = M_{\tilde{T}}$ or (b) $M_Q = 2 M_{\tilde{T}}$ (right).} \label{figs:xsec}
\end{figure}

In order to assess the potential effects of the operators~\eqref{eq:bbg}, we implement the theoretical framework described in Section~\ref{sec:fields} into {\sc FeynRules} and generate the corresponding UFO model~\cite{Christensen:2009jx,Degrande:2011ua,Alloul:2013bka}. In this way, the {\sc MadGraph5\_aMC@NLO}~\cite{Alwall:2014hca} platform can be used to handle signal event generation and cross section calculations. In Figure~\ref{figs:xsec}, we present cross sections for the single production of the top and bottom partners $T_1$ and $B_1$ via their chromomagnetic interactions at a center-of-mass energy $\sqrt{s} = 14$ TeV. We focus on the two processes
\be
  p p \to B_1  + {\rm H.c.}\qquad\text{and}\qquad
  p p \to T_1 \bar t + {\rm H.c.}\, ,
\ee
    and we additionally compare the obtained cross section predictions with pair-production estimates (related to the QCD-induced process $pp\to T_1\bar T_1$). In our calculations, leading-order matrix elements are convolved with the leading-order set of NNPDF2.3 parton densities~\cite{Ball:2013hta}, we have set $\epsilon_L = \epsilon_R = 0.7$ and $\Lambda = 5$ TeV; we have considered two scenarios in which (a) $M_Q = M_{\tilde{T}}$ and (b) $M_Q = 2M_{\tilde{T}}$ respectively. Moreover, we have independently verified our findings using {\sc CalcHep} \cite{Semenov:2014rea} with a  implementation of the model generated by the {\sc LanHep} package~\cite{Belyaev:2012qa}.

In the case of $M_Q = M_{\tilde{T}}$, the top and bottom partners $T_1$ and $B_1$ have similar masses and $B_1$ single production is the dominant production channel. The corresponding cross section is a factor of a few larger than the one corresponding to the $pp\to T_1\bar t$ process. This cross section difference originates from a twofold interplay. While the $bg$ luminosity (relevant for $B_1$ production) is smaller than the $gg$ luminosity (relevant for $T_1 \bar t$ production), that is balanced by the fact that $B_1$ production is a $2\to 1$ process and $T_1 \bar{t}$ production is a $2\to 2$ process. In the case of the $M_Q = 2M_{\tilde{T}}$ scenario, the bottom partner $B_1$ is much heavier than the $T_1$ state so that the $2\to 1$ process features a significant phase-space suppression; the dominant vector-like quark production mode then becomes $T_1\bar{t}$ associated production.

In Figure~\ref{figs:xsec}, we additionally compare the two single production total rates with the QCD-induced production cross section of a pair of top partners $T_1 \bar T_1$. Whilst such a production mechanism is first suppressed due to a reduced available phase space, relative to the single production channels, the cross section dependence on the vector-like quark mass further exhibits a steeply falling behavior with increasing values of $m_{T_1}$. Such behavior is typical of the production of a pair of heavy colored particles through strong interactions. In the rest of this section, we consider the dominant single production modes and we study two novel channels to search for $T_1$ and $B_1$ partners at the LHC. To demonstrate their relevance, we estimate their projected sensitivities at the HL-LHC.

\subsection{Single Production of the $B_1$ Partner}\label{sec:ppB}
\begin{figure}
\centering
\begin{subfigure}{0.45\textwidth}
\includegraphics[width=\textwidth]{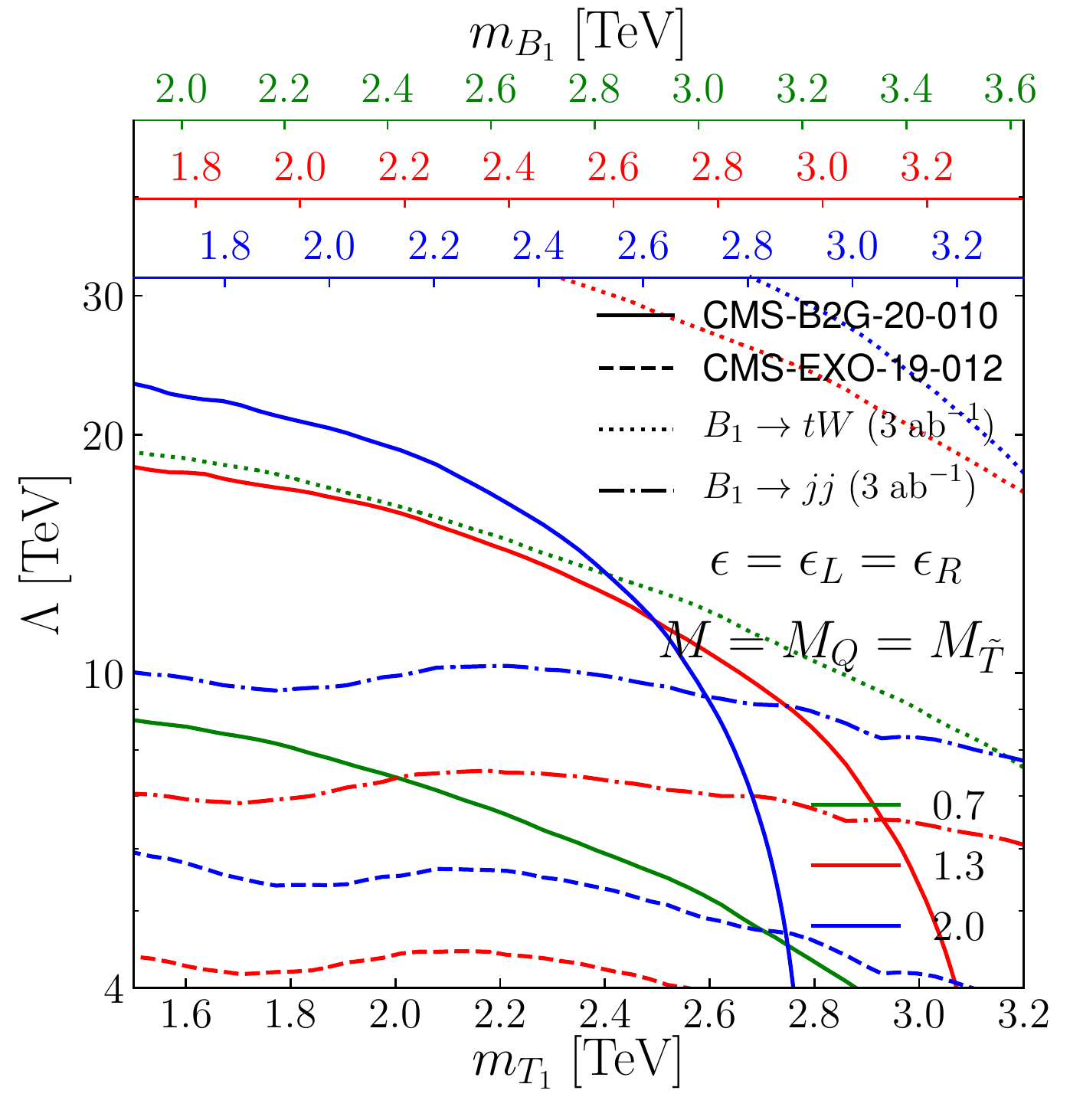}
\caption{}
\end{subfigure}
\begin{subfigure}{0.45\textwidth}
\includegraphics[width=\textwidth]{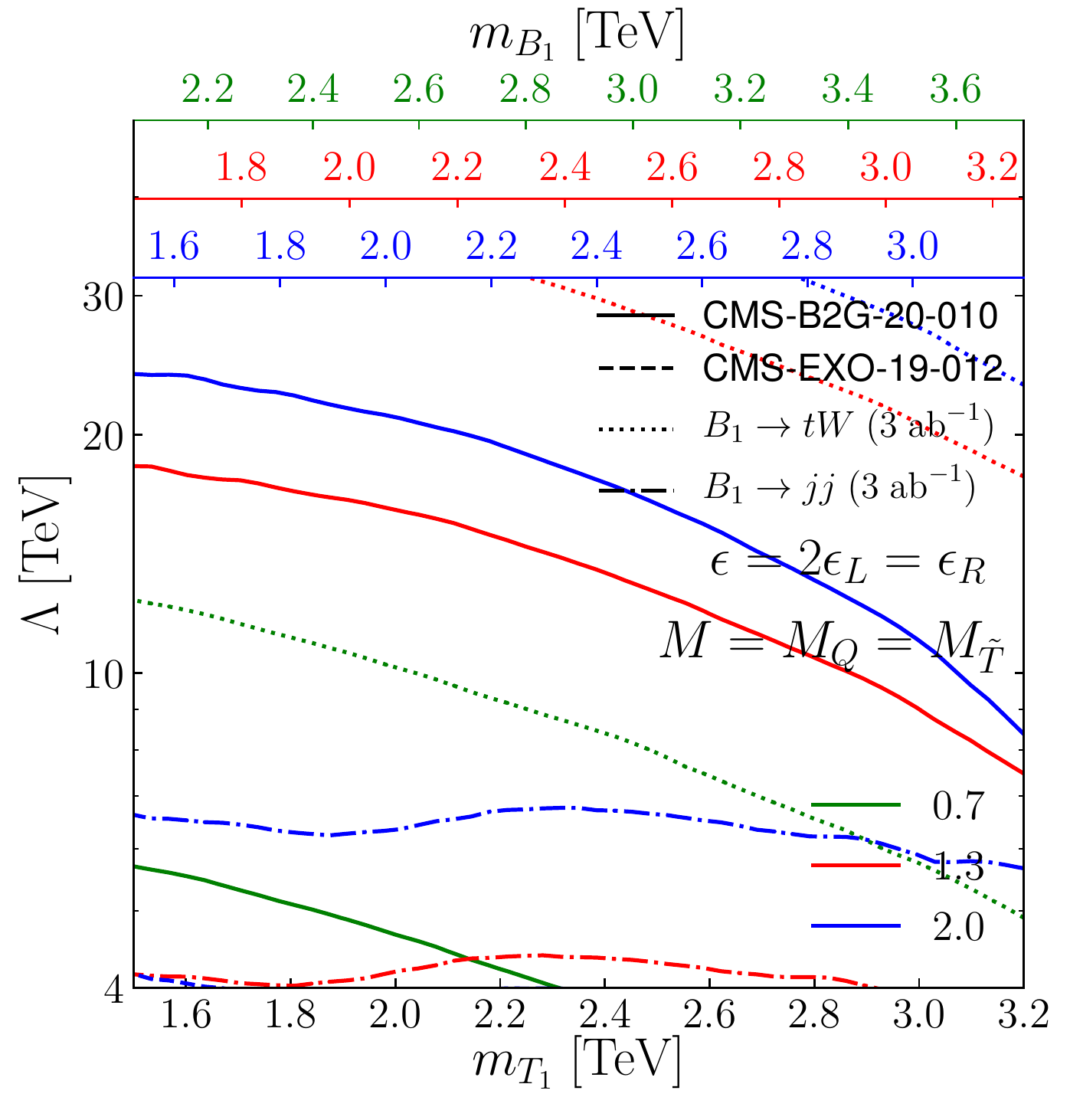}
\caption{}
\end{subfigure}\\
\begin{subfigure}{0.45\textwidth}
\includegraphics[width=\textwidth]{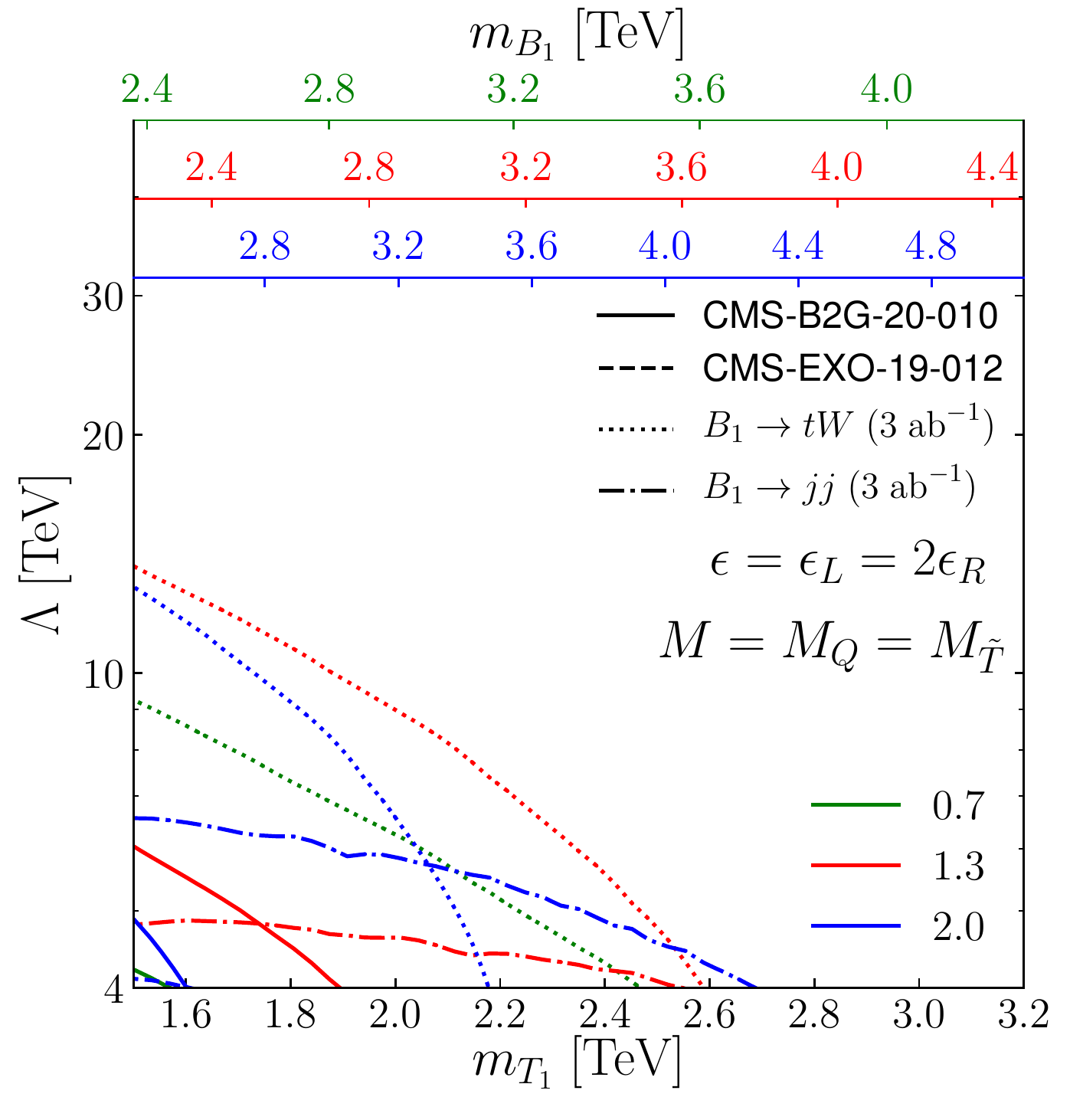}
\caption{}
\end{subfigure}
\caption{Upper bounds on the model parameter space extracted from the CMS searches for the production of an excited $b$-quark decaying into either a $tW$ system (\cite{CMS:2021ptb}; solid) or a $bg$ system (\cite{CMS:2019gwf}; dashed). We provide results in the $(m_{T_1}, \Lambda)$ mass plane for $M_Q = M_{\tilde{T}}$, the corresponding $B_1$ masses being given by the upper $x$-axes for (a)  $\epsilon\equiv\epsilon_L=\epsilon_R$, (b) $\epsilon\equiv2\epsilon_L=\epsilon_R$, and (c) $\epsilon\equiv\epsilon_L=2\epsilon_R$ with $\epsilon = 0.7$ (green), 1.3 (red) and 2 (blue). HL-LHC projections are given by the dotted and dash-dotted lines.}\label{figs:cms_B1}
\end{figure}

In this section, we consider the production of a single bottom partner $B_1$ in proton-proton collisions,
\begin{equation}
p p \rightarrow B_1 \ \   + \ \   {\rm H.c.}
\end{equation}
Depending on the mass difference between the $B_1$ and the $T_1$ states and on the strength of the chromomagnetic interactions ({\it i.e.}~on $\Lambda$), the produced $B_1$ quark can dominantly decay into either a $tW$ system, a $bg$ system or a $T_1W$ system. The first two decay channels have been considered as primary targets for excited quark searches at the LHC by the CMS collaboration~\cite{CMS:2021ptb,CMS:2019gwf}. In Figure~\ref{figs:cms_B1}, we recast the bounds originating from those experimental studies to constrain our model parameter space. We present our results simultaneously in the $(m_{T_1}, \Lambda)$ plane (lower $x$-axis) and in the $(m_{B_1},\Lambda)$ plane (upper $x$-axes), for benchmark scenarios satisfying $M_Q = M_{\tilde{T}}$ and (a) $\epsilon\equiv\epsilon_L=\epsilon_R$ , $\epsilon\equiv2\epsilon_L=\epsilon_R$, or (c) $\epsilon\equiv\epsilon_L=2\epsilon_R$. We consider different values for the off-diagonal mixing parameter $\epsilon$ and focus on a configuration in which the mixing parameter is small ($\epsilon=0.7$, green), moderate ($\epsilon=1.3$; red) and large ($\epsilon=2$; blue).

When the $m_{B_1} - m_{T_1}$ mass splitting is small, as in the scenarios considered in Figure~\ref{figs:cms_B1}(a), the $B_1$ quark decays into $tW$ and $bg$ systems with an appreciable rate. This consequently leads to substantial sensitivity of the existing searches to partner quarks with masses lying in the $1.5-3$~TeV range. In the case of $\epsilon\equiv2\epsilon_L=\epsilon_R$, $B_1$ quark predominantly decays into $tW$, resulting in even strong bounds from the $tW$ search channel. Once the mass spliting becomes large enough, as considered in the scenarios depicted in Figure~\ref{figs:cms_B1}(c), the $tW$ and $bg$ decay channels are relatively suppressed due to the large partial width of the $B_1\rightarrow T_1W$ mode. Current searches therefore turn out to be only able to weakly constrain the model. The situation stays similar after naively extrapolating the bounds to the HL-LHC luminosity of 3~ab$^{-1}$ (through the rescaling of the current reach by the square root of the luminosity). The sensitivity that could be expected by searching for bottom partners through their decays into $tW$ and $bg$ systems will remain weak as long as the $B_1\rightarrow T_1W$ decay channel has enough phase space to proceed, as visible from the dotted and dash-dotted lines shown in the two subfigures.

The above results demonstrate that to fully explore the parameter space, it is important to directly search for the complementary decay mode $B_1 \rightarrow T_1 W$, as the latter is uncovered by the current LHC experimental program. In the rest of this subsection, we thus focus on the process
\begin{equation}
  pp\to B_1\to T_1 W \to (t\ h) W \qquad\text{with}\qquad h\rightarrow b\bar{b},
\label{eq:Bsignal}\end{equation}
where we consider a subsequent $T_1$ decay into a (boosted) top quark and a (boosted) Higgs boson that itself decays into a pair of $b$-jets. Such a final state features a high $b$-jet multiplicity which can be used to efficiently suppress the SM background. In order to avoid dealing with the rejection of the overwhelming QCD multi-jet background, we consider a signature in which the top quark decays hadronically while the $W$ boson originating from the initial $B_1$ decay decays leptonically. In this case, the dominant contributions to the SM background stem from $t\bar{t}+$jets production, which we will take into account to estimate the collider sensitivity to the signal (\ref{eq:Bsignal}).

We make use of the chain of tools described in Section~\ref{sec:pairsingle} to generate leading-order hard-scattering events associated with the process~\eqref{eq:Bsignal}, for a center-of-mass energy $\sqrt{s}=14$~TeV. We then match those events with parton showering as modelled by {\sc Pythia}~8~\cite{Sjostrand:2014zea}, that we also use to simulate hadronization. We next rely on {\sc Delphes}~3~\cite{deFavereau:2013fsa} for the fast simulation of the detector response, using the standard HL-LHC detector parameterization shipped with the program. There, the particle-level clustering of hadrons into jets is performed by means of the anti-$k_T$ algorithm~\cite{Cacciari:2008gp}, as implemented in {\sc FastJet}~\cite{Cacciari:2011ma}. The SM $t\bar{t}$ background is simulated similarly, although we merge event samples relying on matrix elements featuring up to two additional jets, following the MLM-matching scheme~\cite{Mangano:2006rw,Alwall:2008qv} as implemented in {\sc MadGraph5\_aMC@NLO} and with a matching scale set to $Q_{\rm match} = 150$~GeV. In order to acquire better statistics in the relevant part of the phase space, generator-level cuts are implemented so that the transverse momentum $p_T(t)$ of one of the final-state top quarks is large,
\begin{equation}
p_T(t) > 400~{\rm GeV,}
\label{eq:cut_g1}\end{equation}
and the generated $t\bar t$ events exhibit a significant amount of (parton-level) transverse activity
\begin{equation}
p_T(t) + p_T(\bar{t}) + \sum_{\rm jets} p_T(j) > 800~{\rm GeV.}
\label{eq:cut_g2}
\end{equation}

At the reconstructed level, we base our analysis on the boosted topology of the signal, the top quark and the Higgs boson that are issued from the heavy top-partner decays featuring a rather large transverse momentum. We consider as a collection of jets the ensemble of jet candidates clustered with a radius parameter $R = 0.8$ and with a transverse momentum satisfying
\begin{equation}
p_T > 200~{\rm GeV}.
\label{eq:cut_r1}
\end{equation}
Two of those fat jets are identified with a top quark and with a Higgs boson by means of their soft-drop mass $M_{\rm SD}^{\rm top}$ and $M_{\rm SD}^{\rm Higgs}$~\cite{Larkoski:2014wba}, that we respectively impose to satisfy
\begin{equation}
140~{\rm GeV}< M_{\rm SD}^{\rm top} < 210~{\rm GeV} \qquad\text{and}\qquad
90 ~{\rm GeV}< M_{\rm SD}^{\rm Higgs} < 140~{\rm GeV}.
\end{equation}
Moreover, we require that the top-candidate fat jet $t_{\rm had}$ contains a slim $b$-tagged jet $b_{\rm had}$ (clustered with a radius parameter $R=0.2$), the angular distance in the transverse plane between the two jets satisfying 
\begin{equation}
\Delta R(b_{\rm had}, t_{\rm had}) < 0.8,
\label{eq:cut_r3}
\end{equation}
and that the Higgs-candidate fat jet $h$ contains at least two slim $b$-tagged jets $b_{h1}$ and $b_{h2}$ such that 
\begin{equation}
\Delta R(b_{h1}, h) < 0.8 \qquad\text{and}\qquad \Delta R(b_{h2}, h) < 0.8\, .
\label{eq:cut_r9}\end{equation}

We then enforce the events to exhibit a small amount of missing energy (that is expected to originate from the leptonic $W$-boson decay),
\begin{equation}
\slashed{E}_T > 20~{\rm GeV,}
\label{eq:cut_r4}
\end{equation}
and to contain exactly one isolated lepton. Lepton isolation is defined through an isolation variable $I_{\rm mini}$ such that
\begin{equation}
I_{\rm mini} < 0.1,
\label{eq:cut_r5}
\end{equation}
where $I_{\rm mini}$ represents the ratio of the sum of the $p_T$ of all objects lying in a cone of radius ${\cal R}$ centered on the lepton, to the lepton transverse momentum $p_T^\ell$. In this definition, the radius of the isolation cone $\cal R$ is given by
\begin{equation}
  {\cal R} = \frac{10~{\rm GeV}}{{\rm min}({\rm max}(p_T^\ell, 50~{\rm GeV}),200~{\rm GeV})}.
\end{equation}
The three-momentum of the invisible neutrino is reconstructed by imposing that the invariant mass of the $\ell\nu$ system is compatible with the mass of the $W$ boson. We obtain
\begin{equation}
\aligned
p_T^\nu =& \slashed{E}_T, \\
p_L^\nu =& \frac{1}{2(p^\ell_T)^2}\left[(m_W^2+2\vec{p}_T^\ell\cdot \vec{\slashed{E}}_T)p_L^\ell\pm |\vec{p}_\ell|\sqrt{(m_W^2+2\vec{p}_T^\ell\cdot \vec{\slashed{E}}_T)^2-4(p^\ell_T)^2\slashed{E}_T^2}\right],
\endaligned
\end{equation}
in which we choose, for a given benchmark scenario, the solution that minimizes the quantity
\begin{equation}
|m(T_1 W) - m_{B_1}|.
\end{equation}
In this notation, $m(T_1 W)$ stands for the invariant mass of the reconstructed $T_1W$ system.

In order to optimize the analysis and improve its significance, we exploit the gradient boosted decision tree (BDT) method~\cite{Friedman:2001wbq}. We make use of its implementation in the {\sc XGBoost} toolkit~\cite{Chen:2016btl} that offers a fast training speed together with a good accuracy~\cite{10.5555/2996850.2996854}. It includes a novel BDT algorithm dedicated to the handling of sparse data that is in particular useful in our case, as signal and background do not fully populate the event space. The algorithm relies on a set of additive optimizations (or constraints) computed from given variables to classify each event as a signal or as a background event. At each stage of the training process, gradient boosting modifies the existing constraints in order to reduce the amount of classification errors, until no further improvement can be made.

We select, as a set of input variables to the BDT, several properties of the reconstructed hadronic top quark $t_{\rm had}$, leptonic $W$ boson $W_{\rm lep}$, and Higgs boson $h$. We consider their transverse momentum and pseudo-rapidity ($p_T(i)$ and $\eta(i)$ with $i\in [t_{\rm had}, W_{\rm lep}, h]$), the invariant mass $m(i, j)$ of any system made of two of these objects and their angular distance $\Delta R(i,j)$ in the transverse plane (with $i\neq j\in[t_{\rm had}, W_{\rm lep}, h]$). We moreover include the total transverse activity $S_T$ in an event (defined as the scalar sum of the transverse momenta of all reconstructed jets and leptons), the reconstructed transverse activity $S_{T, \rm reco}$ defined by
\begin{equation}
S_{T, \rm reco} = p_T(t_{\rm had}) + p_T(W_{\rm lep}) + p_T(h)\, ,
\end{equation}
the transverse momentum of all reconstructed slim $b$-jets ($p_T(i)$ with $i \in [b_{\rm had}, b_{h1}, b_{h2}]$), the angular distance $\Delta R(t_{\rm had}, b_{\rm had})$ between the $b$-jet associated with the top quark decay and the reconstructed $t_{\rm had}$ system, as well as the invariant mass $m(t_{\rm had}, W_{\rm lep}, h)$ of the system made of the three key objects under consideration. The entire set of BDT inputs is thus given by
\be
  \left\{\bsp
    &p_T(t_{\rm had}),\ p_T(W_{\rm lep}),\ p_T(h),\
     p_T(b_{\rm had}),\ p_T(b_{h1}),\ p_T(b_{h2}),\
     \eta(t_{\rm had}),\ \eta(W_{\rm lep}),\ \eta_T(h),\\
    &\Delta R(t_{\rm had}, W_{\rm lep}),\  \Delta R(t_{\rm had}, h),\ \Delta R(W_{\rm lep}, h),\
     \Delta R(t_{\rm had}, b_{\rm had}),\\
    &m(t_{\rm had}, W_{\rm lep}),\  m(t_{\rm had}, h),\ m(W_{\rm lep}, h),
     m(t_{\rm had}, W_{\rm lep}, h),\
     S_T,\ S_{T, {\rm reco}}
  \esp\right\}\, .
\ee

\begin{figure}
\centering
\begin{subfigure}{0.45\textwidth}
\includegraphics[width=\textwidth]{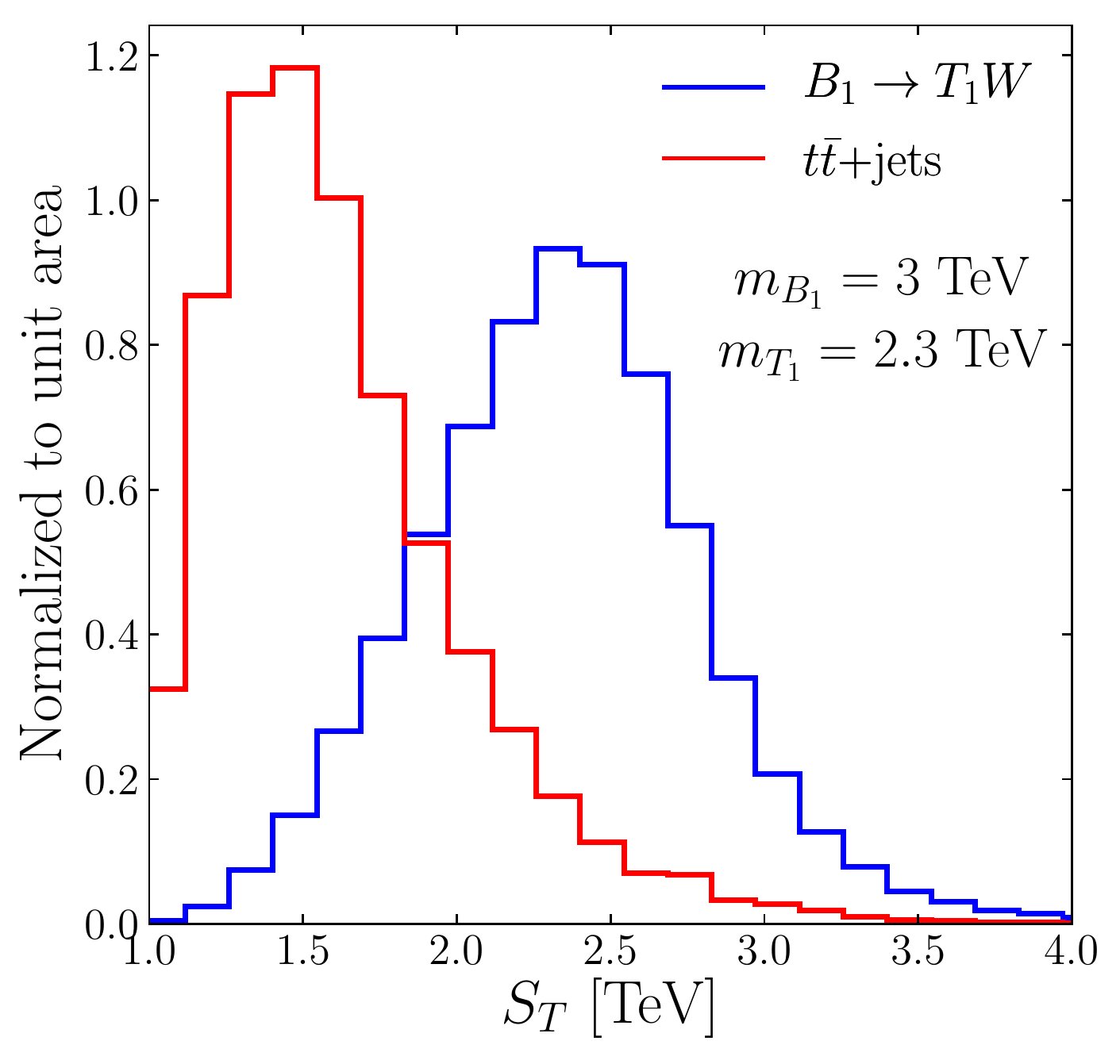}
\caption{}
\end{subfigure}
\begin{subfigure}{0.45\textwidth}
\includegraphics[width=\textwidth]{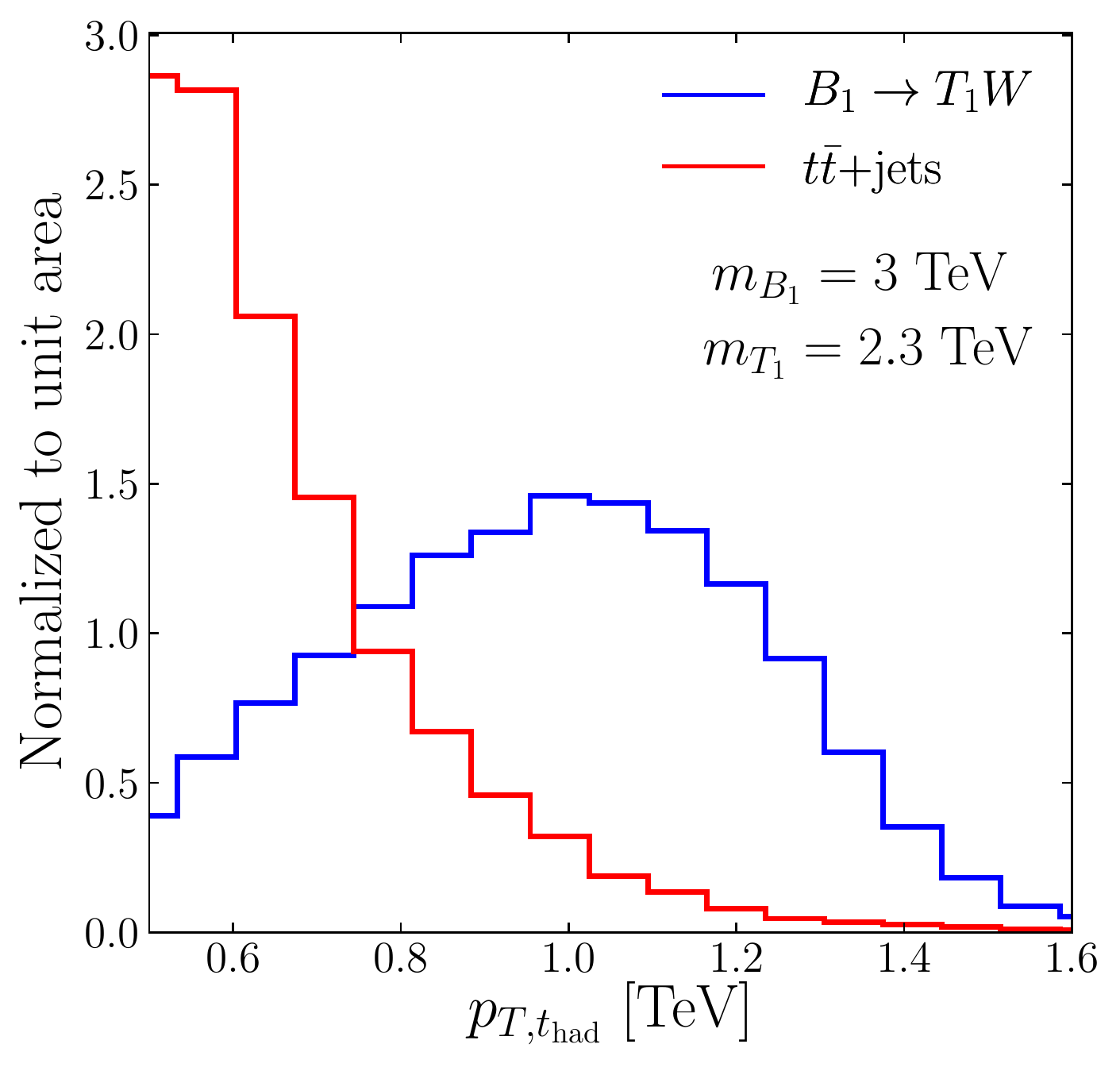}
\caption{}
\end{subfigure}
\\
\begin{subfigure}{0.45\textwidth}
\includegraphics[width=\textwidth]{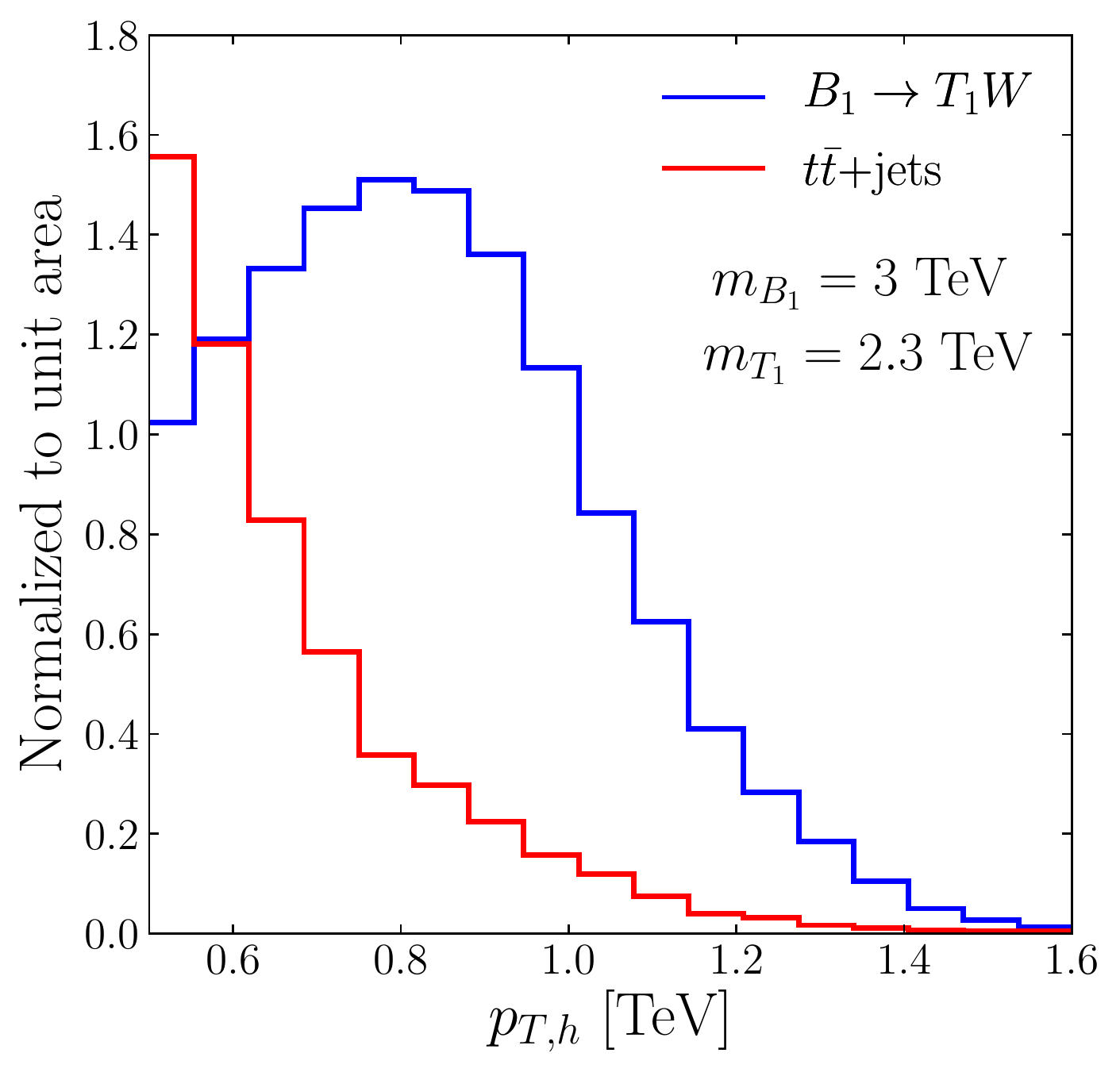}
\caption{}
\end{subfigure}
\caption{Normalized distributions of the total transverse energy in the event $S_T$ (a), of the transverse momentum of the boosted hadronically-decaying top quark $p_T(t_{\rm had})$ (b) and of the transverse momentum of the boosted Higgs candidate $p_T(h)$ (c). We present predictions for the top-antitop SM background (red) and for a $B$-quark signal (blue) for $m_{T_1} = 2.3$~TeV and $m_{B_1} = 3$~TeV, the other model parameters being irrelevant.}
\label{figs:dist_1}
\end{figure}

Among all the variables above, a few of of them are particularly important for signal-background discrimination. As signal events feature a significant amount of transverse activity, the $S_T$ variable is expected to be associated with a very good discriminative power. As shown in Figure~\ref{figs:dist_1}(a) for a signal benchmark scenario in which $m_{B_1} = 3$~TeV and $m_{T_1} = 2.3$~TeV (the other model parameters being irrelevant), the signal distribution peaks at a very large value (around 2~TeV for the benchmark scenario considered), which contrasts with the background that exhibit a much softer spectrum. Whereas the latter features a peak at low $S_T$ value, this peak is artificial and only originates from the generator-level cut~\eqref{eq:cut_g2}. In the panels (b) and (c) of Figure~\ref{figs:dist_1}, we present distributions of the transverse momenta of the boosted hadronic top-quark $t_{\rm had}$ and of the reconstructed Higgs-boson candidate $h$. As the signal spectra are associated with the decay of a heavy resonance, they tend be much harder than the background spectra, and they moreover feature a resonance peak at large values of the observables. In contrast, the background expectations correspond to a steeply-falling behavior, with the distributions thus peaking at very low $p_T$ values.

\begin{figure}
\centering
\begin{subfigure}{0.45\textwidth}
\includegraphics[width=\textwidth]{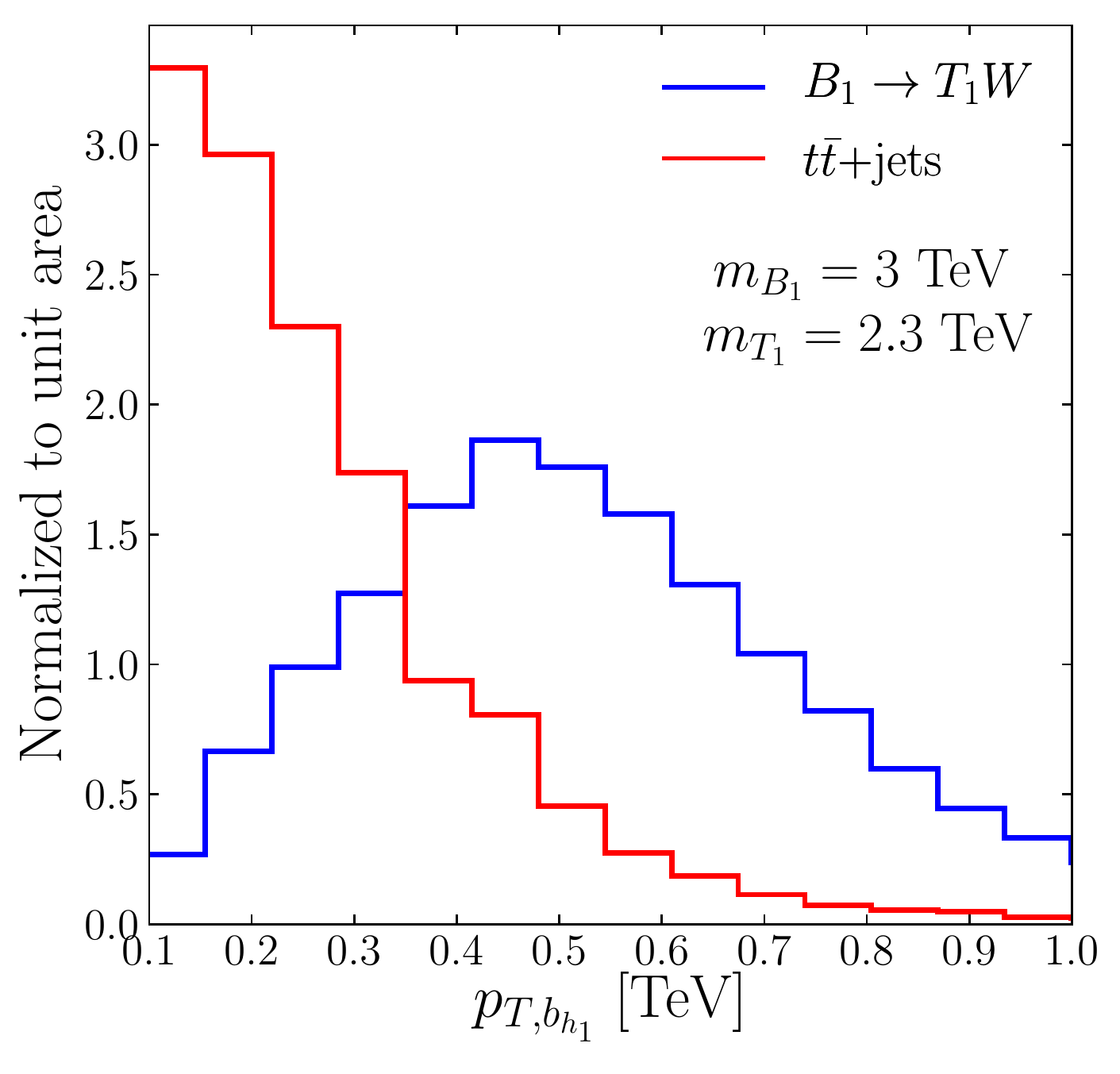}
\caption{}
\end{subfigure}
\begin{subfigure}{0.45\textwidth}
\includegraphics[width=\textwidth]{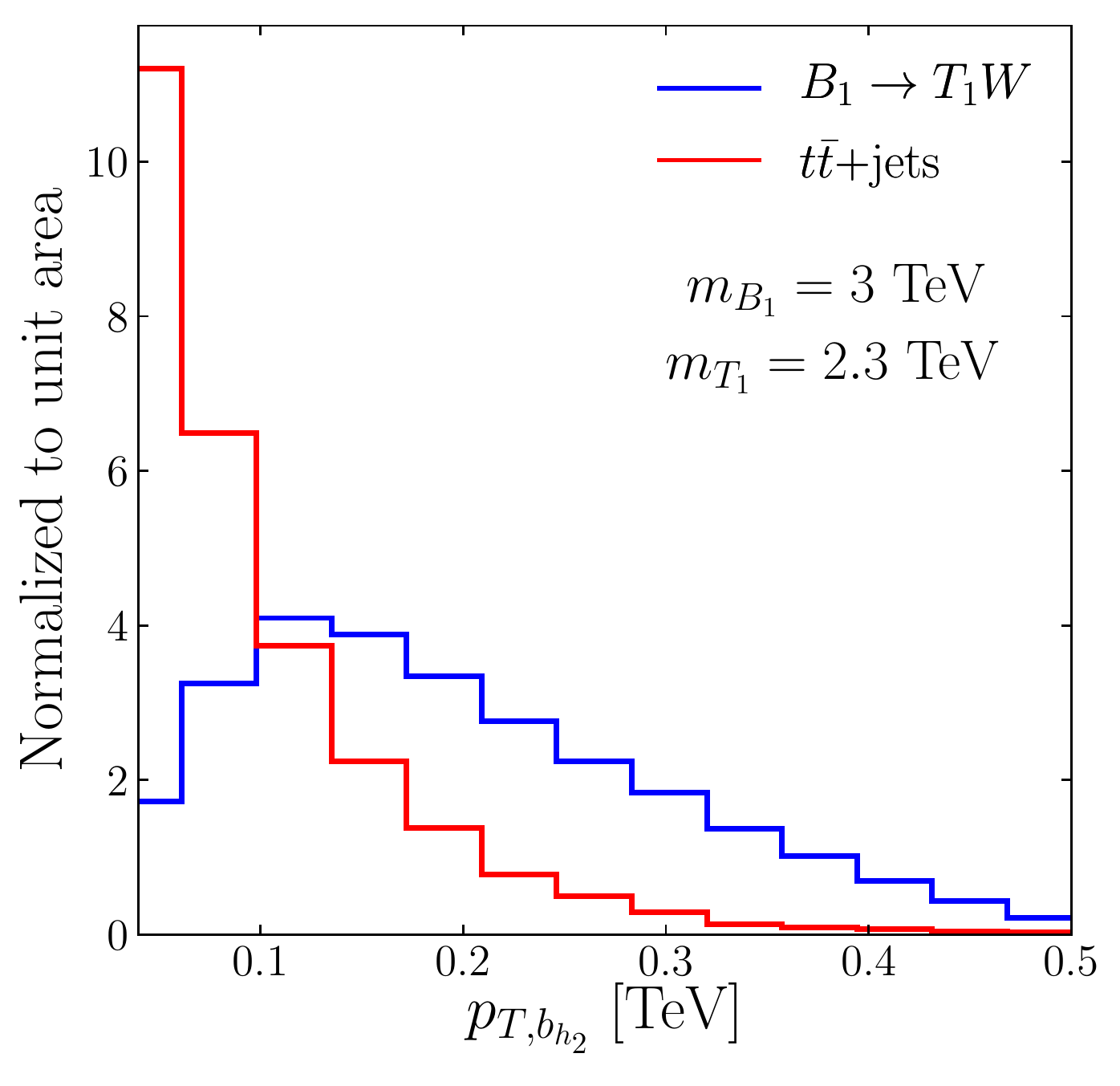}
\caption{}
\end{subfigure}
\caption{Same as in Figure~\ref{figs:dist_1}, but for the transverse-momentum distributions associated with the leading (a) and next-to-leading (b) slim $b$-jet, these two $b$-jets being compatible with a $b\bar b$ decay of a Higgs-boson candidate.}
\label{figs:dist_2}
\end{figure}
Other key observables consist of the $p_T$ spectra of the two slim $b$-jets issuing from the decay of the boosted Higgs boson. These $b$-jets inherit the hardness of the reconstructed Higgs candidate, whose transverse-momentum distribution has been shown in Figure~\ref{figs:dist_1}(c), so that we could potentially expect them to be good discriminators. The $p_T(b_{h1})$ and $p_T(b_{h2})$ distributions, that we present in Figure~\ref{figs:dist_2}, are indeed found to be prime observables to reject the background. The bulk of the background events feature much smaller $p_T$ values of less than 100~GeV, whilst the signal distributions exhibit a peak at larger $p_T$ values. As the Higgs boson present in the signal case is highly boosted, the two $b$-jets significantly overlap in the detector. This impacts their reconstruction so that for many events, they are not perfectly separated. We thus end up with two $p_T$ distributions that are quite asymmetric. While the leading $b$-jet is usually very energetic, the second one carries a much smaller amount of transverse momentum.

\begin{figure}
\centering
\begin{subfigure}{0.45\textwidth}
\includegraphics[width=\textwidth]{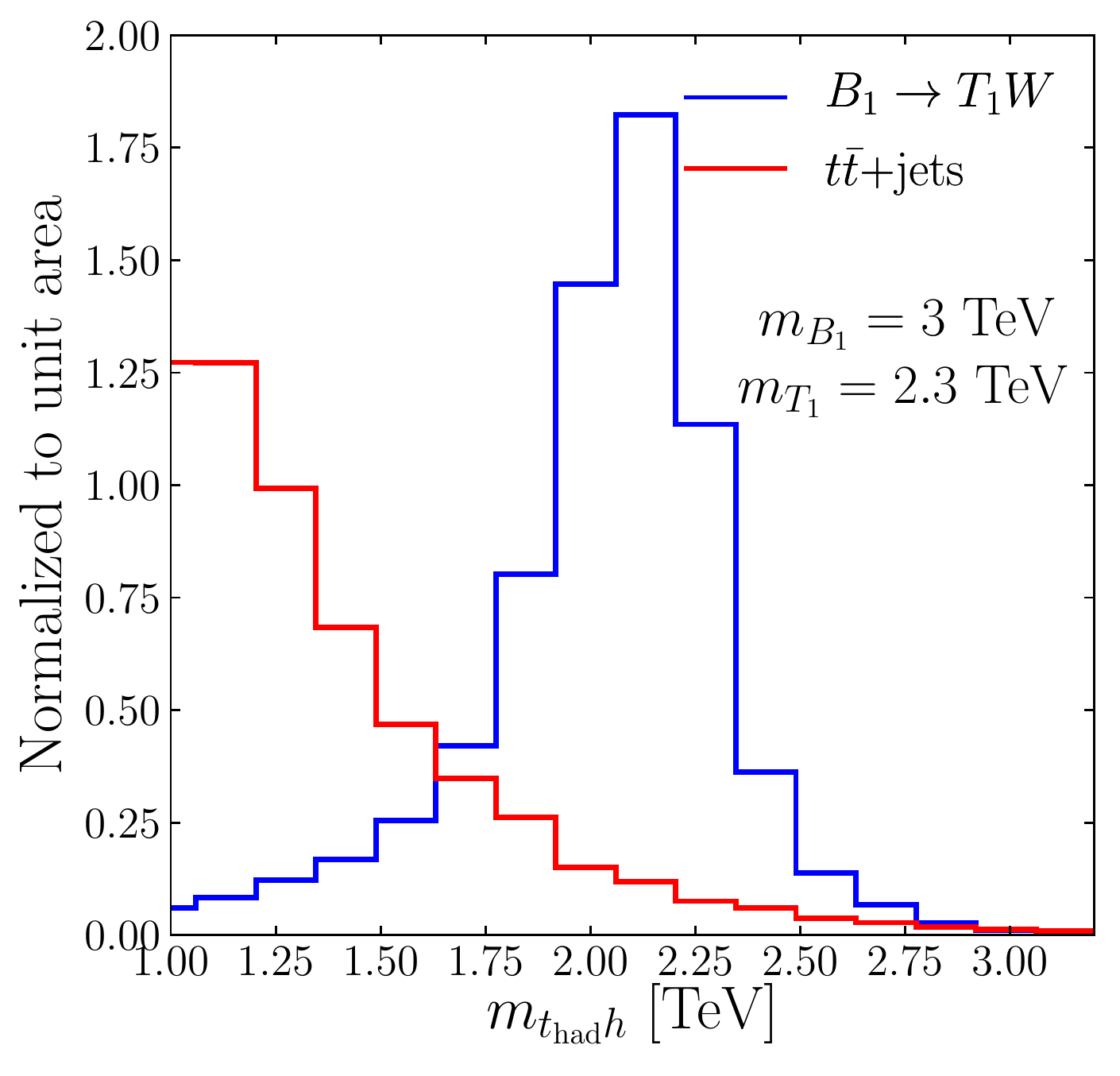}
\caption{}
\end{subfigure}
\begin{subfigure}{0.45\textwidth}
\includegraphics[width=\textwidth]{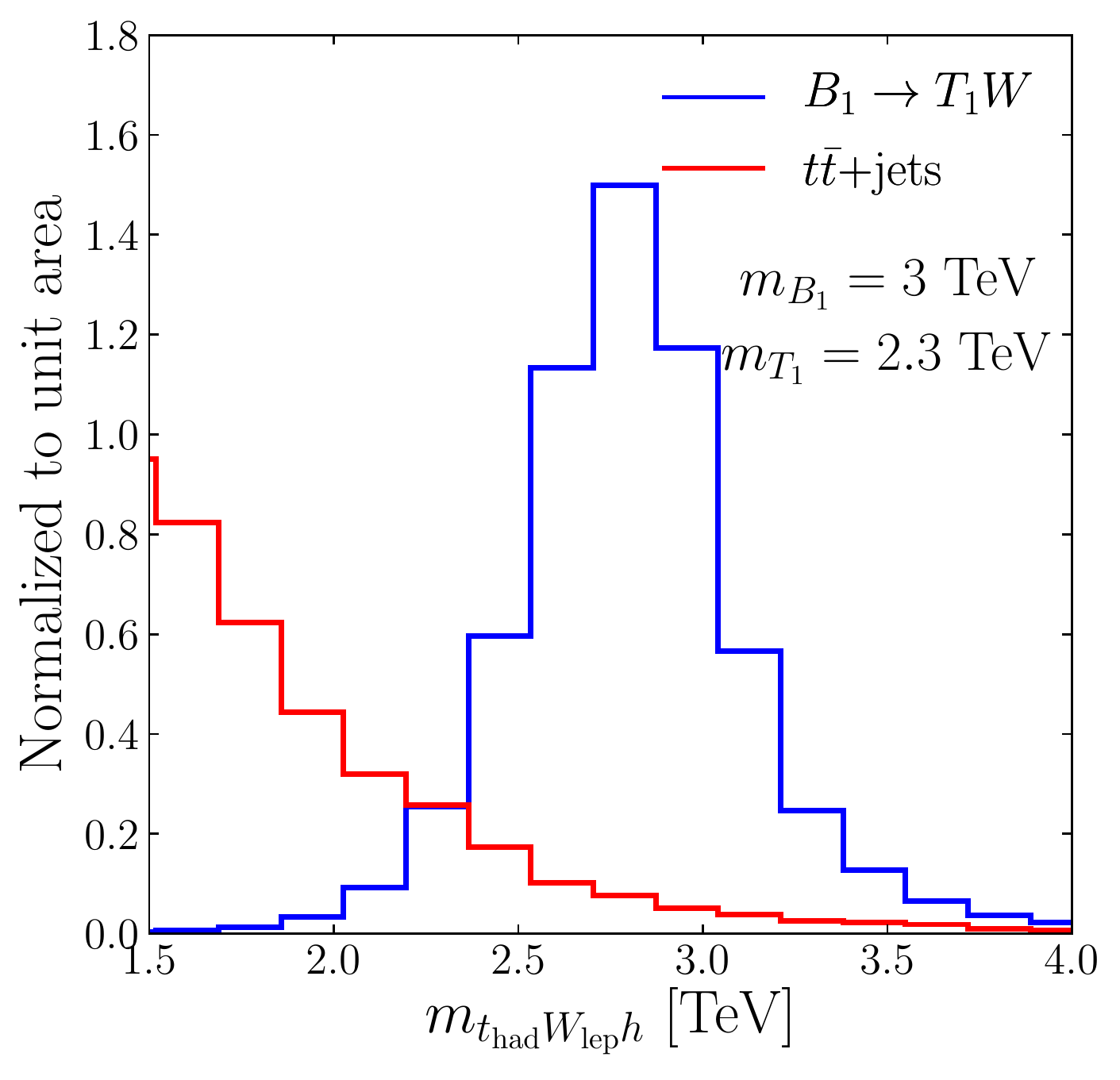}
\caption{}
\end{subfigure}
\caption{Same as in Figure~\ref{figs:dist_1}, but for the distributions in (a) the invariant mass $m(t_{\rm had}, h)$ of the $t_{\rm had}$-Higgs system (associated with a $T_1$ decay), and (b) in the invariant mass $m(t_{\rm had}, W_{\rm lep}, h)$ of the $t_{\rm had}$-$W_{\rm lep}$-Higgs system (associated with a $B_1$ decay).}
\label{figs:dist_3}
\end{figure}

The most powerful observables on which the BDT strategy relies on are two more inclusive variables, namely the invariant mass of the two reconstructed resonances. We present in Figure~\ref{figs:dist_3} the distributions of (a) the invariant mass $m(t_{\rm had}, h)$ of the reconstructed $T_1$-quark, as well as (b) the invariant mass of the reconstructed $B_1$-quark, $m(t_{\rm had}, W_{\rm lep}, h)$. The signal distributions are found to present a peak centered on the true heavy quark masses, although detector effects make those peaks much broader than the physical widths of the $B_1$ and $T_1$ quarks. In contrast, the background distributions feature once again a steeply-falling behaviour, with very few events ending up in the tail of the distribution and with the bulk of them lying at low values of the observables.

We optimize the selection cut on the returned BDT scores to get the best estimate of the HL-LHC sensitivity to the considered $B_1$ signal. The results are provided and detailed in Section~\ref{sec:sensitivity}, together with those stemming from the $pp\to T_1 t$ channel studied in Section~\ref{sec:tT1}.

\subsection{Single Production of the $T_1$ Partner}\label{sec:tT1}
The $s$-channel-resonant production of a $B_1$ quark is a striking signature of the model that we consider, and it nicely complements the searches currently on-going at the LHC. However, it becomes less relevant when the $B_1$ quark is significantly heavier than the $T_1$ quark by virtue of the cross-section dependence on the $B_1$ mass. As such a configuration is realized in a large region of the parameter space, we explore in this subsection a complementary channel in which single $T_1$ production directly proceeds via the considered chromomagnetic moment, through the process
\begin{equation}
p p \rightarrow T_1\bar{t}\ \   + \ \   {\rm H.c.}
\end{equation}
We have seen in Section~\ref{sec:pairsingle} that such a channel could become the dominant production mode of the signal for vector-like-quark spectra in which the lighter $T_1$ and $B_1$ states are quite split. As in Section~\ref{sec:ppB}, we consider a $T_1$ decay into a boosted Higgs boson and a (boosted) top quark, and we then focus on a Higgs boson decaying into a $b\bar b$ system. The full process thus reads
\begin{equation}
 p p \to T_1 t \rightarrow (t\ h)\ t \qquad\text{with}\qquad h\rightarrow b\bar{b}\, .
\end{equation}
Once again, such a process features a high $b$-jet multiplicity, that we could use as a handle on the background. In order to evade the overwhelming QCD multi-jet background, we consider a signal topology in which the boosted top quark ({\it i.e.}\ the top quark that originates from the top-partner decay) decays hadronically while the spectator top quark ({\it i.e.}\ the top quark that is produced in association with the $T_1$ state) decays semi-leptonically. The dominant contributions to the SM background arise from $t\bar{t}+$jets production, which we consider as the sole background in our analysis. Any other potential background components, such $tW$, $t\bar{t}h$ or diboson production (in association with extra hard jets), are indeed expected to be subleading after the selection cuts of our analysis. They are therefore neglected.

The preselection of the analysis undertaken in this section is very similar to the one of Section~\ref{sec:ppB}. We reconstruct the boosted top quark and the boosted Higgs boson in the same way, and impose that the final state of the selected events also contains a single isolated lepton (that is this time assumed to originate from the decay of the spectator top quark $t_{\rm lep}$). We additionally require the presence of an extra slim $b$-jet, clustered with a radius parameter $R=0.4$. Such a $b$-jet, that we denote by $b_{\rm lep}$ in the following, is considered as originating from the decay of the spectator top quark $t_{\rm lep}$, and is imposed to be well separated in the transverse plane from the boosted top quark $t_{\rm had}$. We hence enforce that
\begin{equation}
\Delta R(b_{\rm lep}, t_{\rm had}) > 0.8\, .
\end{equation}
In order to account for the degeneracy due to the high $b$-jet multiplicity, we identifiy as $b_{\rm lep}$ the $b$-jet that forms with the final-state lepton a two-body system whose invariant mass $m(b, \ell)$ satisfies
\begin{equation}
m(b, \ell) < 173~{\rm GeV,}
\label{eq:cut_r7}
\end{equation}
and that can be paired with the missing momentum to yield a system whose invariant mass $m(b, \ell, \nu)$ is compatible with the mass of the top quark. In practice, the latter requirement is achieved by reconstructing the neutrino momentum through a kinematic fit, as in Section~\ref{sec:ppB}, and by minimizing the quantity
\begin{equation}
\Big|m(b, \ell, \nu) - m_t\Big|.
\end{equation}

We then once again base our analysis on a classification relying on boosted decision trees. As a set of input variables, we follow the strategy outlined in the previous section and choose the following observables,
\be\left\{\bsp
    &p_T(t_{\rm had}),\ p_T(t_{\rm lep}),\ p_T(h),\
     p_T(b_{\rm had}),\ p_T(b_{\rm lep}),\ p_T(b_{h1}),\ p_T(b_{h2}),\\
    &\eta(t_{\rm had}),\ \eta(t_{\rm lep}),\ \eta_T(h),\\
    &\Delta R(t_{\rm had}, t_{\rm lep}),\  \Delta R(t_{\rm had}, h),\ \Delta R(t_{\rm lep}, h),\
     \Delta R(t_{\rm had}, b_{\rm had}),\\
    &m(t_{\rm had}, t_{\rm lep}),\  m(t_{\rm had}, h),\ m(t_{\rm lep}, h),
     m(b_{\rm lep}, \ell, \nu),\
     m(t_{\rm had}, t_{\rm lep}, h),\
     S_T,\ S_{T, {\rm reco}}
\esp\right\}\, ,\ee
where the reconstructed activity in the event is defined by
\be
  S_{T, \rm reco} = p_T (t_{\rm had}) + p_T(t_{\rm lep}) + p_T(h)\, .
\ee
We then optimize the selection on the returned BDT scores to get the best possible HL-LHC sensitivity to the considered $T_1t$ signal. The results are analysed and presented in the next section.

\subsection{Sensitivity of the HL-LHC to Single Vector-Like Quark Production through their Chromomagnetic Moments}
\label{sec:sensitivity}

\begin{figure}[htbp]
\centering
\begin{subfigure}{0.45\textwidth}
\includegraphics[width=\textwidth]{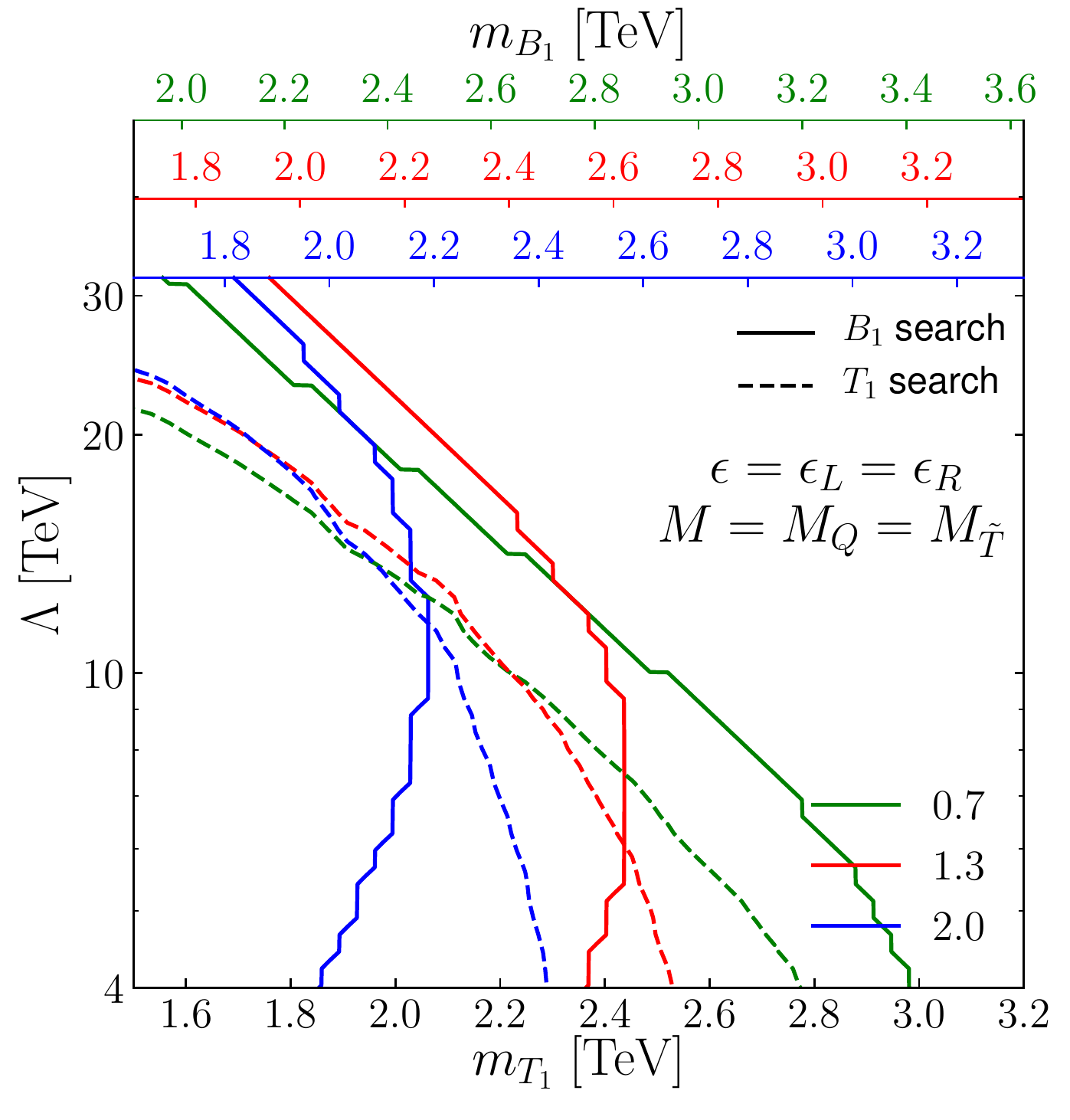}
\caption{}
\end{subfigure}%
\begin{subfigure}{0.48\textwidth}
\includegraphics[width=\textwidth]{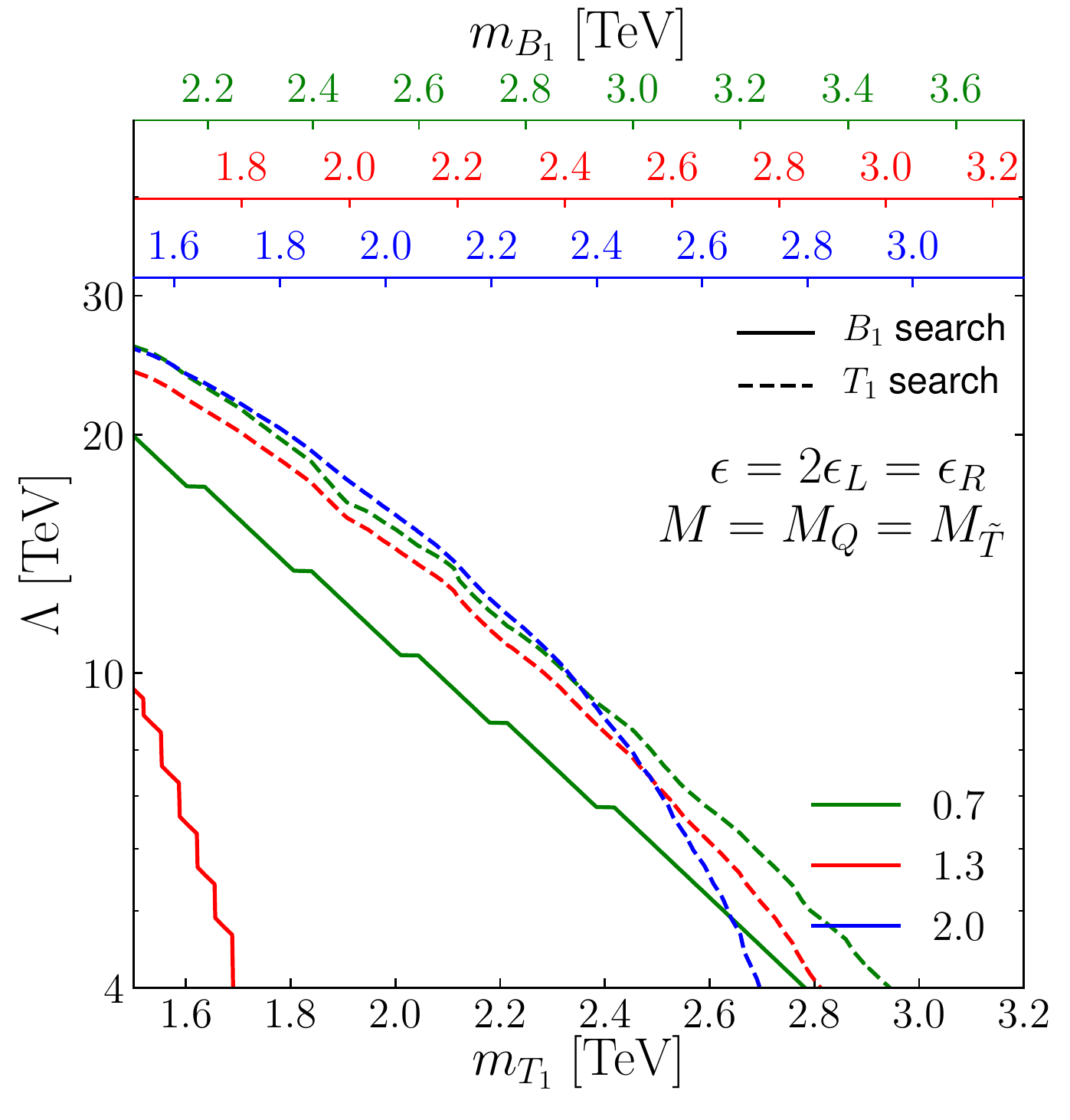}
\caption{}
\end{subfigure}
\\
\begin{subfigure}{0.45\textwidth}
\includegraphics[width=\textwidth]{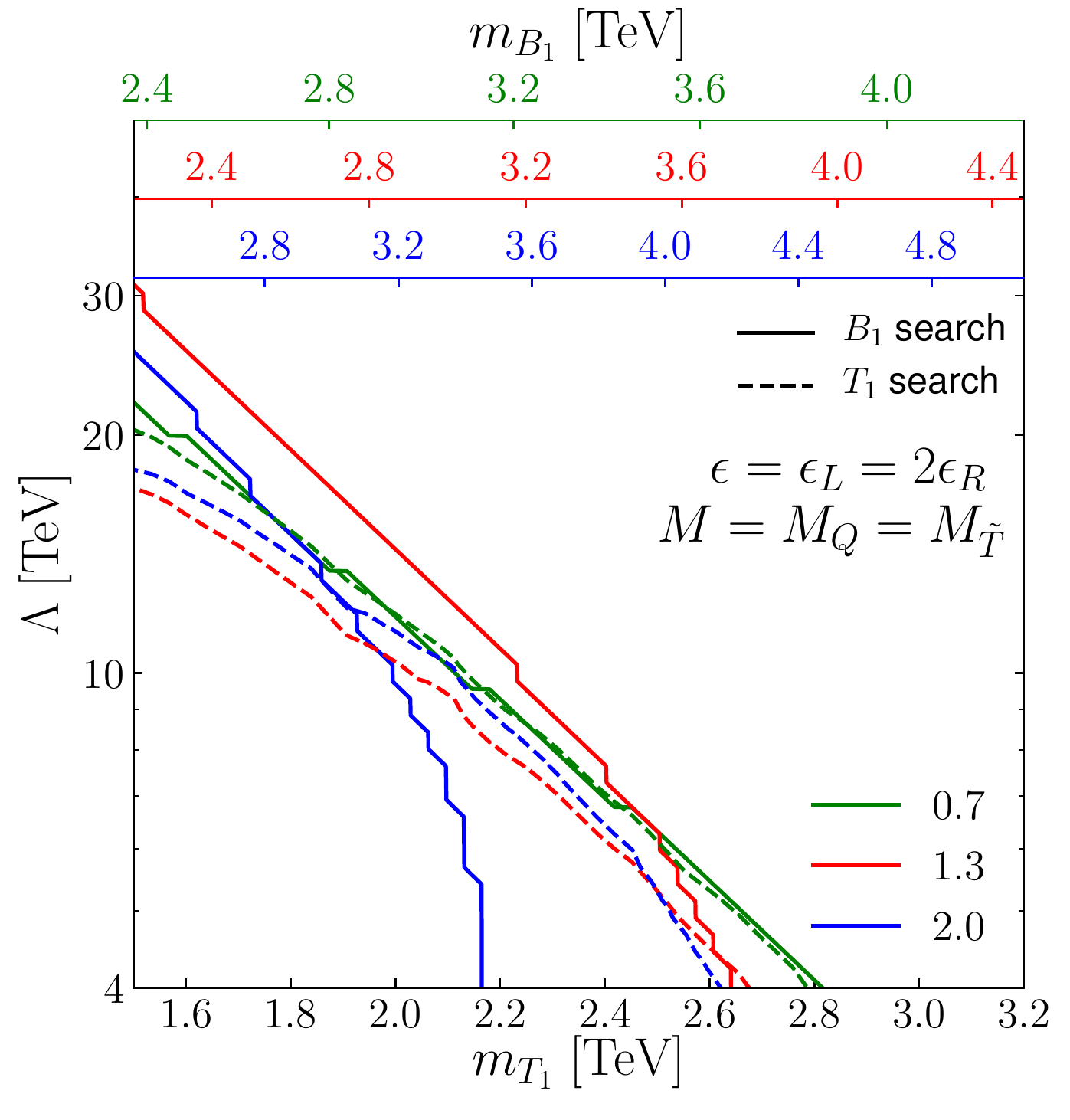}
\caption{}
\end{subfigure}
\caption{HL-LHC sensitivity to the $pp\rightarrow B_1\rightarrow T_1W$ (solid) and $pp\rightarrow T_1 t$ (dashed) signals, shown as two-dimensional 95\% C.L.\ exclusion contours in the $(m_{T_1}, \Lambda)$ plane, with the excluded regions being toward the lower left in each diagram. We  study scenarios in which $M_Q=M_{\tilde{T}}$ and $\epsilon = 0.7$ (green), $1.3$ (red), and $2$ (blue), the mixing being defined as $\epsilon \equiv \epsilon_L = \epsilon_R$ (a), $\epsilon \equiv 2\epsilon_L = \epsilon_R$ (b) and $\epsilon \equiv \epsilon_L = 2\epsilon_R$ (c). The corresponding $B_1$-quark masses $m_{B_1}$ are shown at the top of the figures.}
\label{figs:final_1}
\end{figure}

\begin{figure}[htbp]
	\centering
	\begin{subfigure}{0.48\textwidth}
		\includegraphics[width=\textwidth]{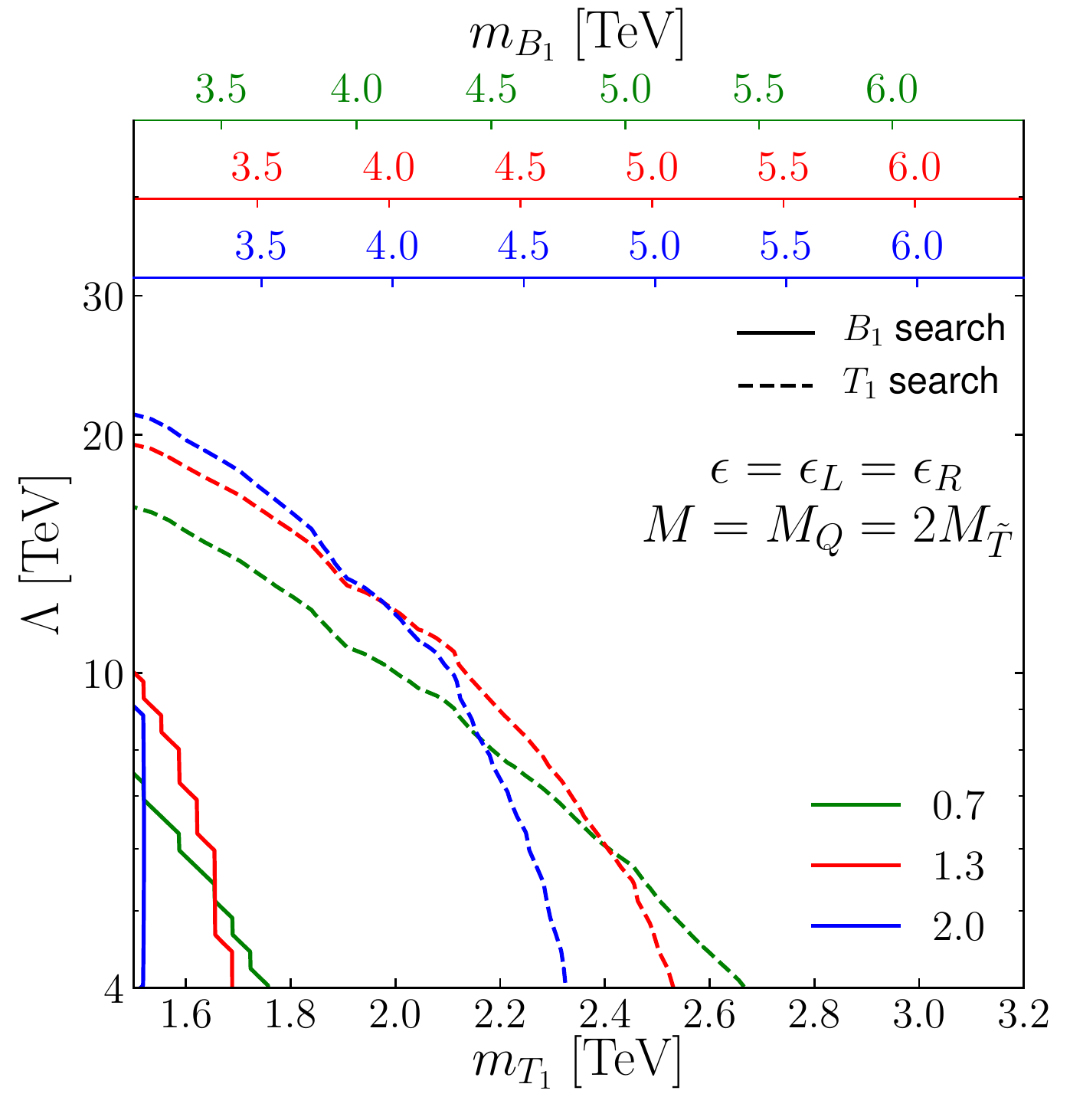}
		\caption{}
	\end{subfigure}%
	\begin{subfigure}{0.45\textwidth}
		\centering
		\includegraphics[width=\textwidth]{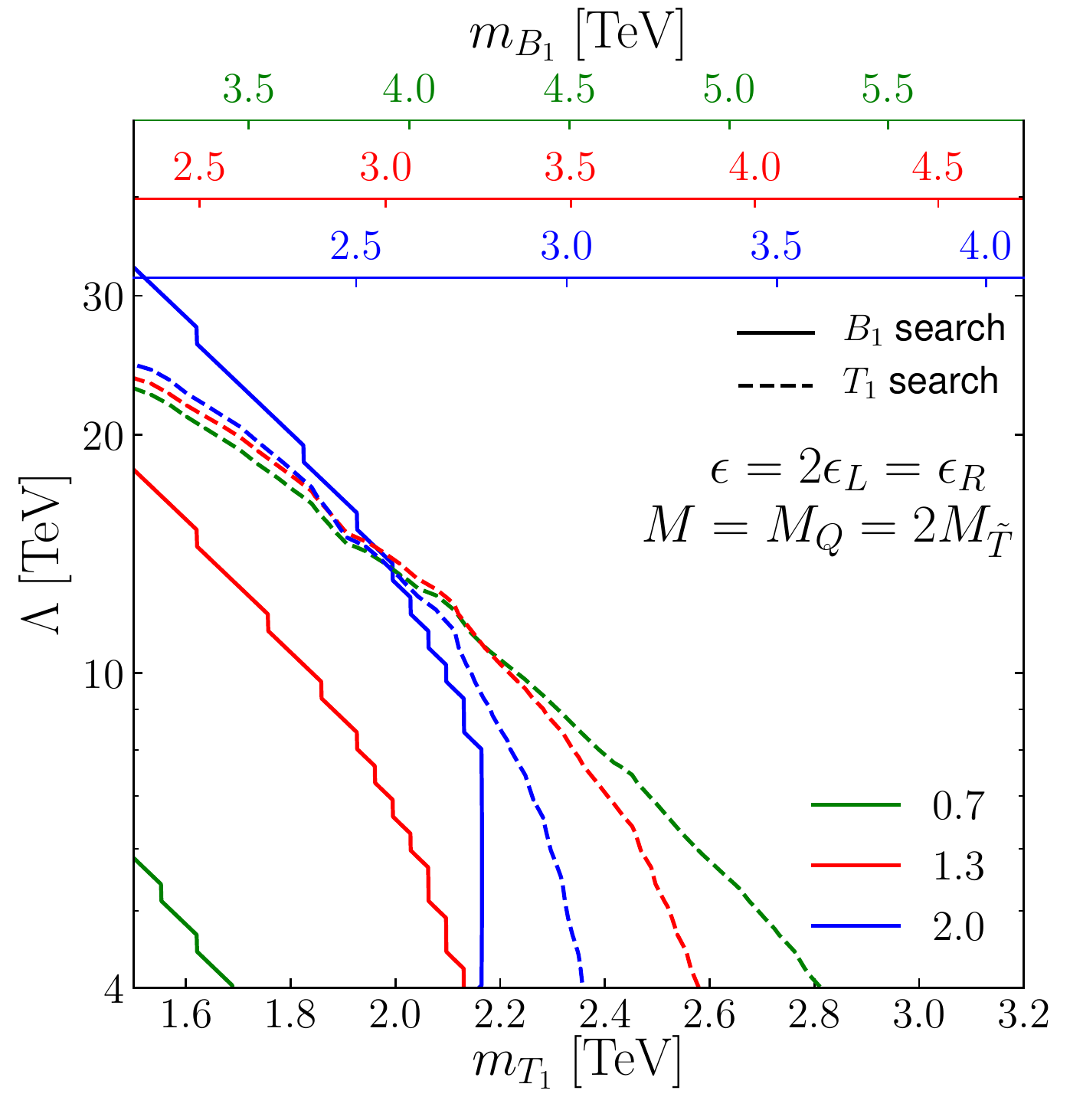}
		\caption{}
	\end{subfigure}
	\begin{subfigure}{0.45\textwidth}
		\centering
		\includegraphics[width=\textwidth]{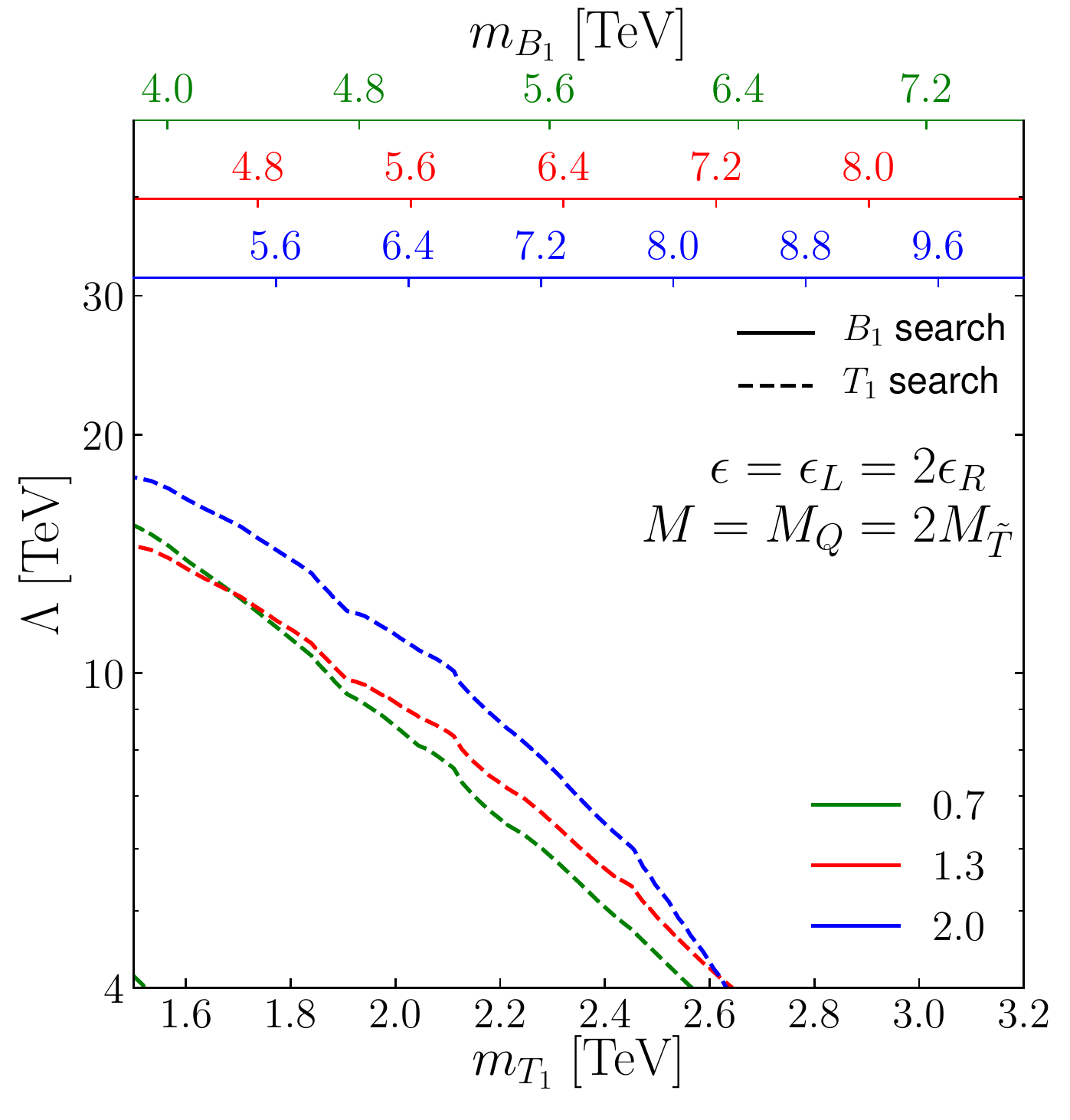}
		\caption{}
	\end{subfigure}
	\caption{Same as Figure~\ref{figs:final_1} but for $M_Q=2 M_{\tilde{T}}$.}
	\label{figs:final_2}
\end{figure}

To estimate quantitatively the sensitivity of the HL-LHC to the two signals considered, we define their statistical significance $Z$ by using~\cite{Cousins:2008zz,Cowan:2010js}
\begin{equation}
Z = \sqrt{2\left((S+B)\ln\frac{S+B}{B} - S\right)}\, ,
\end{equation}
where $S$ and $B$ are the numbers of events for signal and background after all selection cuts respectively. The selection includes a cut on the BDT scores, which has been chosen in order to optimize the significance $Z$. Moreover, we require that at least 3 signal events survive the selection ({\it i.e.}\ $S\geq 3$). The resulting two-dimensional 95\% confidence level (C.L.) contours are presented in the $(m_{T_1}, \Lambda)$ plane in Figures~\ref{figs:final_1} and \ref{figs:final_2} for various scenarios (the excluded regions are at the lower left in each diagram). In Figure~\ref{figs:final_1}, we choose that $M_Q = M_{\tilde T}$ whereas in Figure~\ref{figs:final_2}, we have $M_Q = 2 M_{\tilde T}$. The mixing parameters are taken such that (a) $\epsilon \equiv \epsilon_L = \epsilon_R$, (b) $\epsilon \equiv 2\epsilon_L = \epsilon_R$, or (c) $\epsilon \equiv \epsilon_L = 2\epsilon_R$, and we examine configurations in which the mixing is small ($\epsilon=0.7$; green), moderate ($\epsilon=1.3$; red) and larger ($\epsilon=2$; blue). The exclusions associated with the $pp\to B_1$ analysis (section~\ref{sec:ppB}) are then shown as solid lines, and those associated with the $pp\to T_1t$ analysis (section~\ref{sec:tT1}) are depicted through dashed lines. In each figure, we additionally include an upper horizontal axis for each studied $\epsilon$ value, that we use to represent the corresponding $B_1$ mass values.

For $M_Q = M_{\tilde T}$ (Figure~\ref{figs:final_1}), the largest obtained sensitivity for the parameter range studied in terms of heavy quark masses corresponds to scenarios in which $\epsilon = 0.7$ and $\Lambda = 4$~TeV.\footnote{We do not explore values of $\Lambda$ less than 4.0 TeV in order to maintain the reliability of the effective field theory for the partner masses of interest.} In these cases, bottom-quark partners with masses ranging up to $m_{B_1}\in [3.2, 3.6]$~TeV can be probed. When the mixing increases ($\epsilon = 1.3$ or 2) while $\Lambda$ remains small, the expected reaches of the analyses proposed in this work worsen, as the branching ratios related to the $B_1\rightarrow T_1 W$ and $T_1\to th$ decay modes decrease and/or the quark partner spectrum becomes more compressed. This consequently reduces the signal cross sections by several means. On the other hand, decays into other final states become more frequent, so that this loss of sensitivity is compensated by the expectation relative to the already existing searches for $b$-quark partners (as shown in Figure~\ref{figs:cms_B1}). As long as $\Lambda$ is small, the pattern described above turns out to be realised. When $\Lambda$ gets larger, the impact of the chromomagnetic operators is progressively reduced. This leads to a non-trivial interplay between the spectrum, the branching ratios of the new states and their production cross sections, which manifests itself through the complex form of the observed exclusions. 
Comparing  Figure~\ref{figs:final_1} panels (a) and (b) with Figure~\ref{figs:cms_B1}(a) and (b) we see that - for much of the parameter space investigated here - the $pp\rightarrow B_1\rightarrow T_1W$ and $pp\rightarrow T_1 t$  signals discussed here will allow for discovery of the top-quark partner to complement the excited-quark searches for the bottom-partner.

Very importantly, the existing searches are especially insensitive to scenarios in which $\epsilon = \epsilon_L = 2\epsilon_R$, see Figure~\ref{figs:cms_B1}(c). Our results shows that there is actually a quite promising LHC sensitivity to this configuration, allowing for discovery of both top- and bottom-partners provided extra channels such as those proposed in the present study are added to the LHC experimental program. Compositeness scales of $\Lambda = 20-30$~TeV can even be reached for top-partner masses of about 1.5~TeV.

%  {\bf {\color{red} The text does not ever mention the case $\epsilon =  2\epsilon_L = \epsilon_R$ at all; it is shown in figure 9 but not in figure 5.  Surely it deserves some comment?}}

In Figure~\ref{figs:final_2}, we consider scenarios in which $M_Q = 2 M_{\tilde T}$. Consequently, the spectrum is more split and the $m_{B_1}$ mass is much larger than the $m_{T_1}$ mass (by more than a factor of 2). The direct production process $pp\to B_1$ is thus suppressed so that the analysis introduced in Section~\ref{sec:ppB} loses its sensitivity. The expected LHC reach for the models is therefore entirely dictated by the performance of the analysis of the associated production mode $pp\to T_1t$ that we have detailed in Section~\ref{sec:tT1}. The largest obtained sensitivity in terms of heavy quark masses corresponds again to the smallest $\Lambda$ values considered ({\it i.e.}\ $\Lambda= 4$~TeV), top-quark partners with masses ranging up to about $[2.4, 2.8]$~TeV being reachable regardless of the heavy quark mixing parameters. A similar situation as in the $M_Q = M_{\tilde T}$ case is observed for increasing $\epsilon$ values, as well as for increasing $\Lambda$ values. In this configuration, the existing searches have no sensitivity (regardless of the relative sizes of $\epsilon_L$ and $\epsilon_R$), so that the analysis proposed in Section~\ref{sec:tT1} provides a novel (and unique, so far) promising avenue to explore realistic composite models at the LHC.

%%%%%%%%%%%%%%%%%%%%%%%%%%%%%%%%%%%%%%%%%%%%%%%%%%%%

\section{Conclusions}

In this paper we have investigated the potential for the LHC to discover top-quark partner states produced via their chromomagnetic moment interaction, either via single-production of a bottom-quark partner state which subsequently decays to a top-quark partner and a top-quark, or through the direct production of a top-partner in association with a top-quark. These production mechanisms complement the traditional searches which have relied on pair-production of top-quark partner states, or single production of these states through electroweak interactions, in the sense of providing greatly increased sensitivity where the traditional searches are relatively uninformative.

Using a simplified model to describe the interactions in this framework, and focusing on the case in which the top-quark partner decays to a top-quark and a Higgs-boson, we find that partner masses of up to about 3~TeV can be reached during LHC run III or HL-LHC, substantially extending the expected mass reach for these new states. Moreover, in the case where $M_Q = M_{\tilde T}$, the new search based on the chromomagnetic moment induced single production opens up the possibility of constraining the case $\epsilon = \epsilon_L = 2 \epsilon_R$ to which existing searches are insensitive. Similarly, for $M_Q = 2M_{\tilde T}$, while conventional searches have a very limited reach for the partner quarks, the single production modes induced by the chromomagnetic moment allow for a wide swath of parameter space to be explored.

In subsequent work, we plan to extend the analyses presented here both theoretically and phenomenologically. On the theoretical side, we will explore how these results would change in a more realistic model with a custodial symmetry. On the phenomenological side, we will consider the complementary top-quark partner decays to a bottom-quark and a $W$-boson (as well as to a top quark and a gluon), though these modes could potentially suffer from larger backgrounds.

In the meantime, we eagerly await new results from the LHC, and potentially the discovery of partner-states of the top-quark!

\label{sec:concl}

%%%%%%%%%%%%%%%%%%%%%%%%%%%%%%%%%%%
\section*{Acknowledgements}
The authors are grateful to Thomas Flacke for the organization of a focus meeting on Fundamental Composite Dynamics at IBS CTPU (Daejeon, Korea) in 2017, where the discussions that have given rise to this work have been initiated. 
Authors acknowledge the use of the IRIDIS High Performance Computing Facility, and associated support services
at the University of Southampton to complete this work. 
RSC, EHS, and XW were supported, in part, by by the US National Science Foundation under Grant No. PHY-1915147.
AB acknowledges partial  support from the STFC grant ST/L000296/1 and Soton-FAPESP grant.

%%%%%%%%%%%%%%%%%%%%%%%%%%%%%

%%%%%%%%%%%%%%%%%%%%%%%%%%%%%%%%%%%%%%%%%%

\newpage

\bibliographystyle{apsrev4-1.bst}

\bibliography{tstar}{}

\end{document}